\RequirePackage{lineno}
\documentclass[floatfix,twocolumn,prd,showpacs,preprintnumbers,amsmath,nofootinbib,amssymb,noeprint]{revtex4-1}
\usepackage{mathtools, cuted}
\usepackage{lipsum}
\usepackage[utf8]{inputenc}
\usepackage{graphicx}
\usepackage{dsfont}
\usepackage{mathrsfs}
\usepackage{bm}
\usepackage{color}
\usepackage[colorlinks,citecolor=blue,urlcolor=blue,linkcolor=black]{hyperref}
\usepackage{braket}
\usepackage{placeins}
\usepackage{multirow}
\usepackage{hyperref}
\usepackage{float}
\usepackage{booktabs}
\usepackage{soul}
\usepackage{float}

\usepackage{tikz,pgfplots,pgfplotstable}
\usepgfplotslibrary{groupplots,dateplot}
\usetikzlibrary{patterns,shapes.arrows}
\pgfplotsset{compat=newest}
\usepgfplotslibrary{fillbetween}
\usetikzlibrary{external}
\usepackage[nameinlink,capitalize]{cleveref}

\makeatletter
	\def\Ginput@path{{./figures/}}
\makeatother

\def\tf{\tau_{\text{F}}}
\def\Nc{N_{c}}

\def\dA{d_{\text{A}}}

\begin{document}

\title{Sphaleron rate from Euclidean lattice correlators: An exploration}

\author{
   Luis Altenkort$^{\rm 1}$, Alexander M.~Eller$^{\rm 2}$, O. Kaczmarek$^{\rm 1,3}$,
    Lukas Mazur$^{\rm 1}$, Guy D.~Moore$^{\rm 2}$, H.-T.~Shu$^{\rm 1}$
}

\affiliation{
    $^{\rm 1}$Fakult\"at f\"ur Physik, Universit\"at Bielefeld, D-33615 Bielefeld,
    Germany \\
    $^{\rm 2}$Institut f\"ur Kernphysik, Technische Universit\"at Darmstadt\\
Schlossgartenstra{\ss}e 2, D-64289 Darmstadt, Germany \\
    $^{\rm 3}$Key Laboratory of Quark and Lepton Physics (MOE) and Institute of
    Particle Physics, \\
    Central China Normal University, Wuhan 430079, China
}


\begin{abstract}

We show how the sphaleron rate (the Minkowski rate for topological charge
diffusion) can be determined by analytical continuation of the Euclidean
topological-charge-density two-point function, which we investigate on the lattice, using gradient flow to reduce noise and provide improved operators
which more accurately measure topology.
We measure the correlators on large, fine lattices in the quenched approximation at $1.5\,T_c$ with high precision. Based on these data we first perform a continuum extrapolation at fixed physical flow time and then extrapolate the continuum estimates to zero flow time. The extrapolated correlators are then used to study the sphaleron rate by spectral reconstruction based on perturbatively motivated models.

\end{abstract}

\maketitle
\section{Introduction}
\label{sec:intro}
The classical vacuum of QCD is not unique; there are an infinite number of
topologically distinct classical vacua defined by different integer values of
the Chern-Simons number.  Transitions between different
classical vacua are important because such transitions change the
axial quark number, which would otherwise be conserved for
massless quarks and slowly changing for light quarks.  The sphaleron rate is defined as the
mean-squared change in Chern-Simons number per spacetime volume.
It was first considered within QCD because of the role it may play
in electroweak baryogenesis \cite{McLerran:1990de,Giudice:1993bb}.
More recently it has raised interest because magnetic phenomena in
heavy ion collisions might give axial charge density an important role
to play in, for example, the chiral magnetic effect
\cite{Kharzeev:2007jp,Fukushima:2008xe}.  At high temperatures, where
the QCD coupling is weak,
there are semiclassical approaches to determine the sphaleron rate,
but they falter long before one considers physically relevant
temperatures~\cite{Moore:2010jd}.  Therefore we will make a
first attempt to study the sphaleron rate nonperturbatively using
the methods of lattice QCD and spectral reconstruction.

We begin by reviewing the definition and properties of the sphaleron
rate in QCD.  Topology in QCD is determined through the topological
charge density $q(x)$ and its integral, the topological charge $Q$,
\begin{align}
\label{Q}
    Q=\int \mathrm{d}^4x\ q(x),\ \ q(x)=\frac{g^2}{32\pi^{2}}\epsilon_{\mu\nu\rho\sigma}\textrm{Tr}\left\{ F_{\mu\nu}(x)F_{\rho\sigma}(x)\right\},
\end{align}
where $\epsilon_{\mu\nu\rho\sigma}$ is the totally antisymmetric tensor
and $F_{\mu\nu}$ is the field strength tensor.
It can be proven that $Q$ is an integer for smooth vacuum-to-vacuum configurations~\cite{tHooft:1976snw}
and is equal to the number of left-handed zero modes
minus the right-handed zero modes of the massless Dirac operator~\cite{Atiyah:1968mp}.
In Minkowski time, the topological charge density determines the violation
of axial current conservation: neglecting quark masses, the axial current
$J_5^{\mu}=\bar{\psi}\gamma^{\mu}\gamma_5\psi$ obeys
\cite{McLerran:1990de} 
\begin{align}
\label{noncon-current}
    \partial_{\mu}J_5^{\mu}=2 N_fq(x),
\end{align}
where $N_f$ is the number of light quark flavors. The sphaleron rate $\Gamma_{\mathrm{sphal}}$ is defined as the rate of the mean squared change in the topological charge per Minkowski 4-volume, or equivalently the integration of the Wightman correlator of the topological charge density.
It can be reexpressed in terms of the spectral function $\rho(\omega)$ of the topological-charge-density two-point correlator
at zero spatial momentum,
\begin{align}
    \Gamma_{\mathrm{sphal}}= 2T\lim_{\omega\rightarrow 0}\frac{\rho(\omega)}{\omega},
\label{spha_formula}
\end{align}
which in turn is related to the \textit{Euclidean}
topological-charge-density correlation function $G(\tau)$
through an integral relation,
\begin{align}
\label{gqq}
\begin{split}
G(\tau)&=\int \mathrm{d}^3x\ \langle q(\vec{0},0) q(\vec{x},\tau)\rangle\\ &=-\int_0^{\infty}\frac{\mathrm{d}\omega}{\pi}\ \rho(\omega)\frac{\cosh[\omega(1/2T-\tau)]}{\sinh(\omega/2T)}.\\
\end{split}
\end{align}
Here $T$ is the temperature and $\tau$ is the temporal distance between the two charge-density operators, and the unusual minus sign arises because $q$ is a time reversal odd operator.

The sphaleron rate is of interest both in electroweak baryogenesis~\cite{Giudice:1993bb,Moore:2010jd,Kharzeev:2019rsy} and heavy ion collisions~\cite{Voloshin:2010ut,Fukushima:2008xe,Kharzeev:2007jp}. For electroweak interacting matter the sphaleron rate has been well understood and determined using B\"{o}deker's effective theory~\cite{Bodeker:1998hm,Bodeker:1999ud,Bodeker:1999ey,Arnold:1999uy} in the weak-coupling regime.  These innovations allowed for a
nonperturbative semiclassical real-time evaluation on a Minkowski lattice
\cite{Moore:1998zk,Bodeker:1999gx}.
A similar study for the weak-coupling behavior of the SU(3) sphaleron rate is given in \cite{Moore:2010jd},
while for the physically interesting coupling regime a direct evaluation in Minkowski space is impossible.
The results of nonperturbative lattice studies are also rather limited.
A recent lattice QCD study on the SU(3) sphaleron rate can be found in \cite{Kotov:2018vyl}.

In this work we attempt to give a constraint on the sphaleron rate by fitting Euclidean correlators to perturbatively motivated models in different scenarios. Due to high-frequency fluctuations of the gauge fields on the lattice, the correlators considered in this work are noisy, and some noise reduction method is necessary. 
Gradient flow \cite{Luscher:2011bx,Luscher:2010iy,Narayanan:2006rf} is the preferred method because it has a valid definition in terms of continuum field theory and we have a good analytical understanding of how it affects the gauge fields. 
This makes it preferable to traditional approaches such as cooling~\cite{BERG1981475,IWASAKI1983159}, or APE~\cite{Albanese:1987ds}, stout~\cite{Morningstar:2003gk}, or HYP~\cite{Hasenfratz:2001hp} smearing.
Related studies can be found in ~\cite{Alexandrou:2015yba,Alexandrou:2017hqw,Bonati:2014tqa}.
Since its introduction, gradient flow has proven to be useful for a variety of issues in lattice QCD~\cite{Kitazawa:2016dsl,Kitazawa:2017qab,Borsanyi:2012zs, Bazavov:2015yea, DallaBrida:2019wur}. It also provides access to investigate how instantons emerge in the continuum limit of lattice QCD
\cite{Ce:2015qha,Taniguchi:2016tjc,Kotov:2018vyl,Mazur:2020hvt}. In this work the gradient flow is mainly employed as a noise reduction technique. The field smearing nature of the gradient flow enables the creation of smooth gauge configurations on the lattice, from which a well-defined topological charge can be obtained \cite{Luscher:2010iy, Luscher:2010we}.

This study is closely related to our previous work~\cite{Altenkort:2020fgs}, which shares the same implementation of the gradient flow and lattice setup. The strategy for the continuum and flow-time-to-zero extrapolation is also similar.
The subtle differences will be addressed in later sections.

In the following we briefly introduce the lattice setup, including our implementation of the topological charge (density) and gradient flow on the lattice (\autoref{sec:setup}).
In \autoref{sec:roletopology} we examine under what conditions our definition of topological density correctly returns the topology,
and we investigate whether topological freezing will affect our results.
Section~\ref{sec:extrap} is devoted to the continuum extrapolation at fixed flow time and the subsequent extrapolation to zero flow time.
In \autoref{spha} we review the perturbative spectral function and combine it with various models for low-frequency contributions
to extract the sphaleron rate from the lattice data.
We then discuss the results and conclude in \autoref{sec:conclusion}.

\section{Lattice Setup}
\label{sec:setup}

The setup in this work is identical to that of our previous work \cite{Altenkort:2020fgs}.
We perform numerical calculations of SU(3) Yang-Mills theory in 4-dimensional spacetime with periodic boundary conditions for all directions. The configurations are generated on large, fine isotropic lattices in the quenched approximation using the standard Wilson gauge action. To make sure that the configurations are sampled from the thermal equilibrium we first perform 5000 heat-bath sweeps. Afterwards we save one configuration after every 500 combined sweeps, where a combined sweep consists of one heat-bath and four over-relaxation sweeps. We have checked
that this procedure eliminates autocorrelations between saved configurations in all
quantities we consider except for total topology, which we discuss later.

With the aim of providing results in the continuum limit, we measure the topological charge density correlators on five different lattices.
For each lattice the $\beta$ value is tuned according to $r_0T_c=0.7457(45)$~\cite{Francis:2015lha} so that all are at almost the same temperature of 
$1.5\,T_c$.
The scale setting is done via the Sommer parameter $r_0$~\cite{Sommer:1993ce} with parametrization from~\cite{Francis:2015lha} and updated coefficients from~\cite{Burnier:2017bod}. We summarize the key information in \autoref{tab:lattice_setup}.

\begin{table}[h]
    \centering
    \begin{tabular}{ccrcccc}                            
    \hline \hline
    $a$ (fm) & $a^{-1}$ (GeV) & $N_{\sigma}$ & $N_{\tau}$ & $\beta$ & $T/T_{c}$ & \#conf.\tabularnewline
    \hline
    0.0262 & 7.534 & 64 &  16  & 6.8736 &  1.51  & 10000 \tabularnewline
    0.0215 & 9.187 & 80 &  20  & 7.0350 &  1.47  & 10000 \tabularnewline
    0.0178 & 11.11 & 96 &  24  & 7.1920 &  1.48  & 10000 \tabularnewline
    0.0140 & 14.14 & 120 & 30  & 7.3940 &  1.51  & 10000 \tabularnewline
    0.0117 & 16.88 & 144 & 36  & 7.5440 &  1.50  & 10000 \tabularnewline
    \hline \hline
    \end{tabular}
    \caption{
    Lattice spacings, lattice extents, $\beta$ values and statistics of configurations in this work. The lattice spacing $a$ is determined by the Sommer scale (see~\cite{Sommer:1993ce}).}
    \label{tab:lattice_setup}
\end{table}

We use a gluonic definition
of the topological charge density $q(x)$ based on an $a^2$-improved implementation of the field strength tensor. Rather than the
standard ``clover'' sum of four square plaquettes, we use a mixture of square
and $1\times 2$ rectangular plaquettes~\cite{BilsonThompson:2002jk}
\begin{align}
    \label{eq:impF}
    F^{\mathrm{Imp}}_{\mu\nu}(n) = \frac{5}{3}C^{(1,1)}_{\mu\nu}(n)-\frac{1}{3}C^{(1,2)}_{\mu\nu}(n),
\end{align}
where $C^{(m,n)}_{\mu\nu}(n)$ are field strength tensors constructed from the variant clover terms 
made up of plaquette rectangles $W^{m\times n}_{\mu\nu}$, see~\cite{BilsonThompson:2002jk} for details.
This is, however, insufficient to provide a $q(x)$ definition which captures topological information, as it suffers both from large renormalization issues and nontopological noise.
In order to obtain a well-defined definition of $q(x)$ that accurately captures topological information, we need to apply gradient flow to smooth out the highest-frequency fluctuations in the gauge fields.  

The gradient flow introduces a flow-time variable $\tf$ and a gauge field
$B_\mu(\tf,x)$ which is set equal to the physical gauge field
$A_\mu(x)$ at $\tf=0$ and subsequently evolves with increasing
flow time $\tf$ by gradient descent under the Yang-Mills action:
\begin{align}
    \label{boundary}
    B_{\mu}(\tf=0,x)=A_{\mu}(x), \quad
    \frac{\partial{B_{\mu}}}{\partial{\tf}}=D^{B}_{\nu}G^{B}_{\nu\mu}.
\end{align}
The superscript $B$ indicates that $D^B_\nu G^B_{\nu \mu}$ is evaluated using
$B_\mu(\tf)$ rather than $A_\mu$.
We adopt a Symanzik improved discretization of the gradient flow called \textit{Zeuthen flow}~\cite{Ramos:2015baa}, which introduces no new $\mathcal{O}(a^2)$ discretization errors.
The discretized flow equation is integrated in small steps using a third-order Runge-Kutta algorithm with an adaptive step-size.
The same numerical implementation of the gradient flow is also employed in~\cite{Altenkort:2020fgs}. The observables are measured on flow times $\sqrt{8\tf}T\in \lbrace 0, 0.001, \dots,0.2,0.205,\dots,0.3\rbrace$.

Gradient flow can be understood as a modification of the operators used in the measurement, replacing the elementary links with unitarized averages over many paths, an extreme form of the use of ``fat links.''
At lowest perturbative order it is equivalent to replacing the gauge fields with their averages over a Gaussian envelope with width $\sqrt{8\tf}$.
But gradient flow is a nonperturbatively well-defined and gauge invariant procedure.

Using gradient flow to modify our operators 
is a two-edged sword.
As we increase the amount of flow, the elimination of UV fluctuations
makes the correlation functions less noisy.  In addition, the
lattice definition of $q(x)$ is always contaminated by nontopological
high-dimension operators suppressed by powers of the lattice spacing $a$,
and if fluctuations with wave number $k\sim 1/a$ are present, these
operators contaminate our measurement with nontopological effects.
By eliminating such short-distance fluctuations, gradient flow improves
the topological behavior of $q(x)$.  As we will soon see, there
is a lattice spacing dependent flow depth above which
$q(x)$ correctly captures topological information.  Smaller values
cannot be trusted and must be avoided.
But the value of the correlation function is $\tf$-dependent and only the 
$\tf \to 0$ limit corresponds to the desired correlation function.
This limit must be extracted by extrapolation; but as we increase $\tf$,
the difference eventually becomes enough that the finite $\tf$ correlator
is not useful in trying to extrapolate to $\tf \to 0$, as
very coarse lattices do not help us to take the continuum limit.
On dimensional grounds, one expects that, for a correlation function
measured over a separation $\tau$, these corrections will be
dependent on $\tf/\tau^2$.  Only sufficiently small $\tf/\tau^2$
values will be in a useful scaling window where they help determine
the $\tf \to 0$ extrapolation.  For larger values physical information
is lost; indeed, above a certain ratio of $\tf/\tau^2$
the correlation function even has the wrong sign \cite{Eller:2018yje}.
Therefore there is also a $\tau$-dependent upper
limit on available $\tf$ values; Ref.~\cite{Eller:2018yje}
conservatively advises $\sqrt{8\tf} < 0.33 \tau$.

\begin{figure}[t!]
    \centerline{\includegraphics[width=0.5\textwidth]{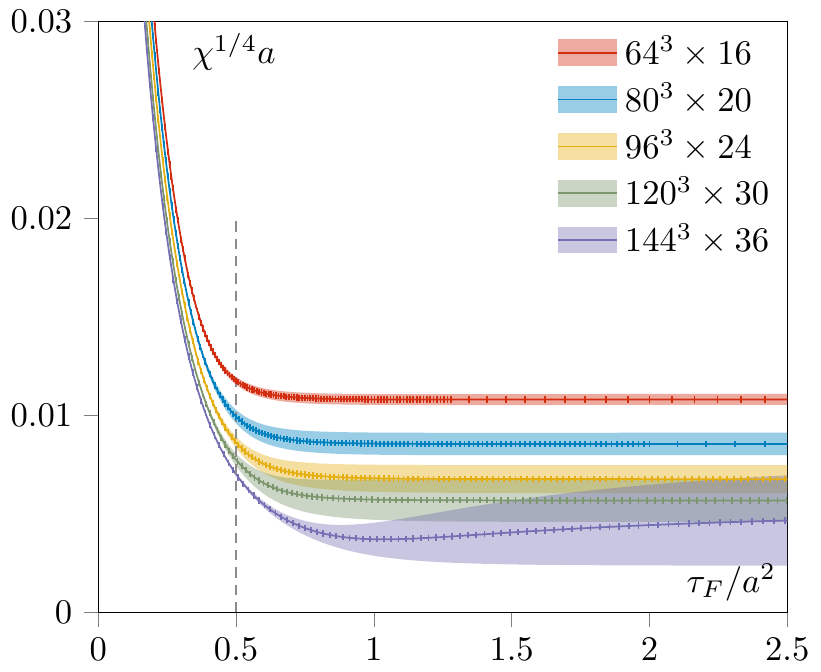}
    }
    \caption{Topological susceptibility measured at different flow times. The vertical line is the flow time which we have chosen as a lower limit to ensure that $q$ properly captures topology.}
    \label{fig:topSuscPlot}
\end{figure}

\section{Role of topology}
\label{sec:roletopology}

\begin{figure}[t!]
    \centerline{\includegraphics[width=0.47\textwidth]{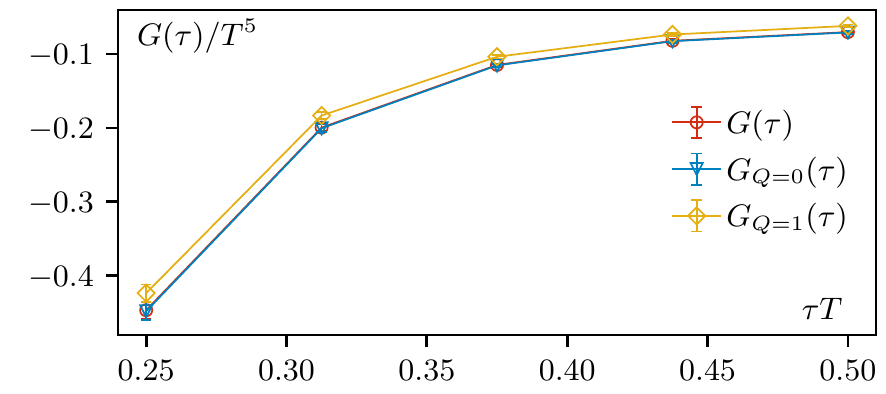}
    }
    \vspace{0.1cm}
    \centerline{\includegraphics[width=0.47\textwidth]{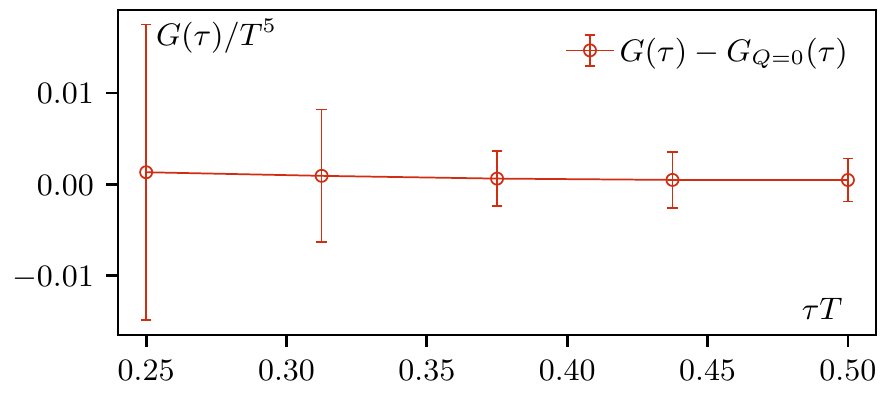}
    }
    \caption{Topological charge density correlator of the $N_\tau=16$ lattice at flow times $\sqrt{8\tau_{F}^{\mathrm{max}}}T=0.5220 \tau T$ (the largest flow times used in the later extrapolation).
    Top: Comparison of the full correlator with the
    correlators of the $Q=0$ and $Q=1$ sectors.
    Bottom: Difference between the full correlator and the
    correlator computed using only the $Q=0$ sector,
    in comparison to the statistical error bars.}  \label{fig:freezingPlot}
\end{figure}

Because the observable $q$ is a topological density, it makes sense to first check what role topology plays in our calculations.  
This means, first, to see what $\tf$ value is necessary before $q$ returns accurate topological information, and second, to check that topological freezing does not significantly affect our results.

If we define $q$ using the field strength from \cref{eq:impF} but without applying any gradient flow, the desired topological density operator mixes with nontopological higher-dimension operators which contribute large noisy contributions to the determined topology. 
This effect is ameliorated as we increase the amount of flow. 
To examine this in detail, we study the topological susceptibility $\chi \equiv \langle Q^2 \rangle / \Omega$ where $\Omega$ is the spacetime volume of the lattice.
We compute this on each lattice at a series of flow times $\tf$. 
Our results, shown in \autoref{fig:topSuscPlot}, indicate that our determined topological susceptibility stabilizes above $\tf = 0.5 a^2$.  For smaller values, the aforementioned artifacts significantly contaminate the determined topological susceptibility. 
We will therefore only rely on data with $\tf \geq 0.5 a^2$ in what follows, so that we can be confident that our operator faithfully represents the topological density.

In analyzing the susceptibility, we found that $Q$ evolves slowly on our finer lattices, a phenomenon called topological freezing.
This is visible in Figure \ref{fig:topSuscPlot} in the much
larger error bars for the susceptibility for the finest lattices.
If the correlation functions we are interested in are sensitive to the
topological sector, then this can cause large autocorrelation problems and
poor statistical power in our results.  
However, it is not clear that correlation functions of the topological density should be especially sensitive to the topological sector.  
After all, the sphaleron rate which we are seeking to compute is not related to the topological susceptibility; at weak coupling the sphaleron rate is completely unrelated to Euclidean topology and instantons \cite{Arnold:1987zg}.
Therefore, we will examine how much the topological charge density correlators are affected by topological sector, and therefore how much of a problem topological freezing may be.
We do this by measuring the $qq$ correlators separately in different topological sectors, and then analyzing how much a misweighting or exclusion of the $Q\neq 0$ sectors might affect our results.
We use the $N_\tau = 16$ lattice for this study because the
topological susceptibility is relatively well determined there.
To improve our determination of the properties of the $Q\neq 0$
sector, we perform a Markov chain in which we prevent $Q=0$ via
an extra accept-reject step, to get a pure sample of $Q \neq 0$ configurations.  We can then compare this with both the $Q=0$ sector,
and the proper mixture of $Q=0$ and $Q\neq 0$ as determined from the topological susceptibility.  The results are shown in
\autoref{fig:freezingPlot}. 
The $Q=1$ sector disagrees with the $Q=0$ sector by more than their
error bars, but nevertheless the difference is so small that,
given the small fraction of $Q=1$ configurations in the full sample,
the difference is smaller than our statistical errors even for
the largest flow time where the errors are the smallest.
This implies that a misweighting of the $Q=1$ sector, even by
a factor of 2 relative to its correct weighting, will change
the correlation functions of interest by less than their
statistical errors and can therefore be ignored.
Since the finer lattices have a still smaller topological
susceptibility and therefore even less $Q=1$ configurations,
we expect that this result holds for those lattices as well.
Therefore we will proceed without considering topological freezing further.

\begin{figure*}[htb]
     \centerline{\includegraphics[width=0.5\textwidth]{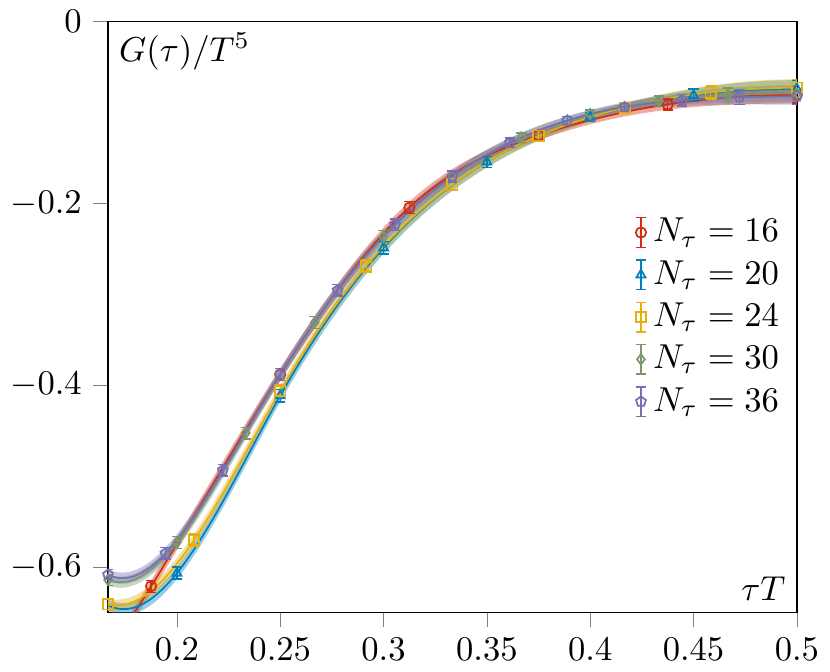}\includegraphics[width=0.475\textwidth]{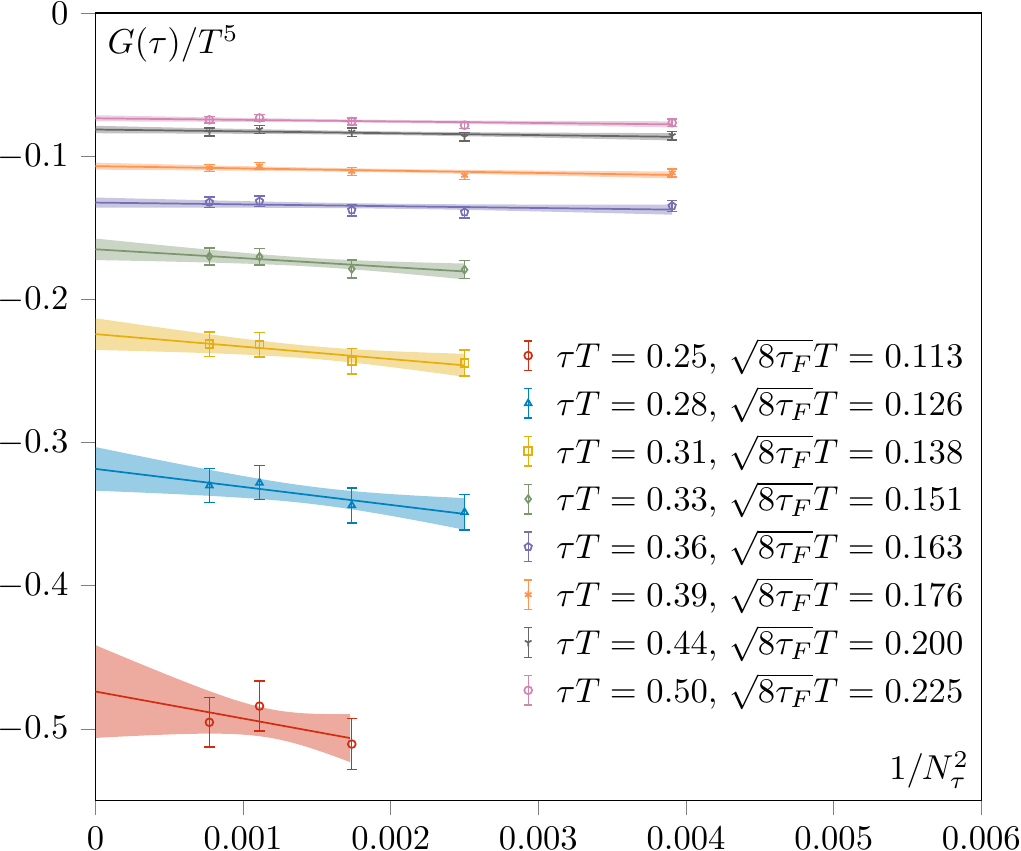}
    }
    \caption{Left: topological charge density correlation function at flow time 
    $\sqrt{8\tau_F}T=0.15$ for all lattice spacings. Error bands represent cubic spline 
    interpolations. 
    Right: continuum extrapolation for some selected separations at flow times halfway between the beginning and the end of each flow range Eq. \eqref{eq:flowlimits}.}  \label{fig:qq_cont_extr}
\end{figure*}

\section{Double extrapolation}
\label{sec:extrap}

Physics resides at zero lattice spacing, and \cref{gqq} only holds for the correlation function at $\tf=0$.
Therefore it is necessary to perform both a continuum and
a flow-time-to-zero extrapolation.  We first perform the continuum
extrapolation at each flow time, and then use these
continuum-limit values of $G_{\tf}(\tau)$ to extrapolate
to $\tf \to 0$.

\subsection{Interpolation}

In order to perform a continuum extrapolation, we first have to interpolate the discretized data of the coarser lattices to the distances of the finest one.
For that we use cubic splines in which the first derivative is constrained to vanish at $\tau T=0.5$ which reflects the symmetry of the correlator on the lattice. 

The data is only interpolated in the range $\tau T \in [0.166,0.5]$.
The correlator at shorter distances is not very helpful in extracting
the low-frequency spectral function, and our scaling window of
useful $\tf$ values closes up at small $\tau T$, so we make no
attempt to study smaller $\tau T$ values.
A set of interpolations at one selected flow time is shown on the left side of \autoref{fig:qq_cont_extr}.

\subsection{Continuum extrapolation}

The continuum extrapolation is performed by linear fits in $1/N_\tau^2$ of the interpolated data at separations of the finest lattice ($N_\tau=36$) using the Ansatz
\begin{align}\label{eq:extrAnsatz}
\frac{G_{\tau,\tau_F}(N_\tau)}{T^5} = m \cdot N_\tau^{-2} + b,
\end{align}
where $b$ is the continuum correlator normalized by $T^5$. To estimate the statistical error, we do bootstrap resampling with $10000$ samples, where each sample consists of $10000$ configurations. The interpolations and extrapolations are done on each sample. 

In the next subsection, we will define minimum and maximum ranges
for $\tf$ to ensure that we are in a scaling window for the $\tf$ extrapolation.
In some cases that window will be in conflict with the condition
$\tf > 0.5 a^2$ obtained in \autoref{sec:roletopology}.
For each $\tau$ value, we will then exclude those lattices where this conflict
occurs from the continuum extrapolation; otherwise we extrapolate
using all available lattices.

The left-hand side of \autoref{fig:qq_cont_extr} shows the topological charge density correlators at
flow time $\sqrt{8\tau_F}T=0.15$ for each lattice spacing. The error bands belong to the
interpolations, and we see that the relative cutoff effects of the lattices are small.
This is also visible on the right-hand side of \autoref{fig:qq_cont_extr}, where we show the
correlator as a function of $1/N_\tau^2$ for some selected separations at flow times chosen from a
flow range which we will define in the next section (see Eq. \eqref{eq:flowlimits}). 
The continuum extrapolated values are within the error bars of the lattice data in every case. 
This is not surprising since the gradient flow produces renormalized operators that are insensitive to lattice-scale fluctuations. 

\begin{figure}[t!]
\centerline{\includegraphics[width=0.46\textwidth]{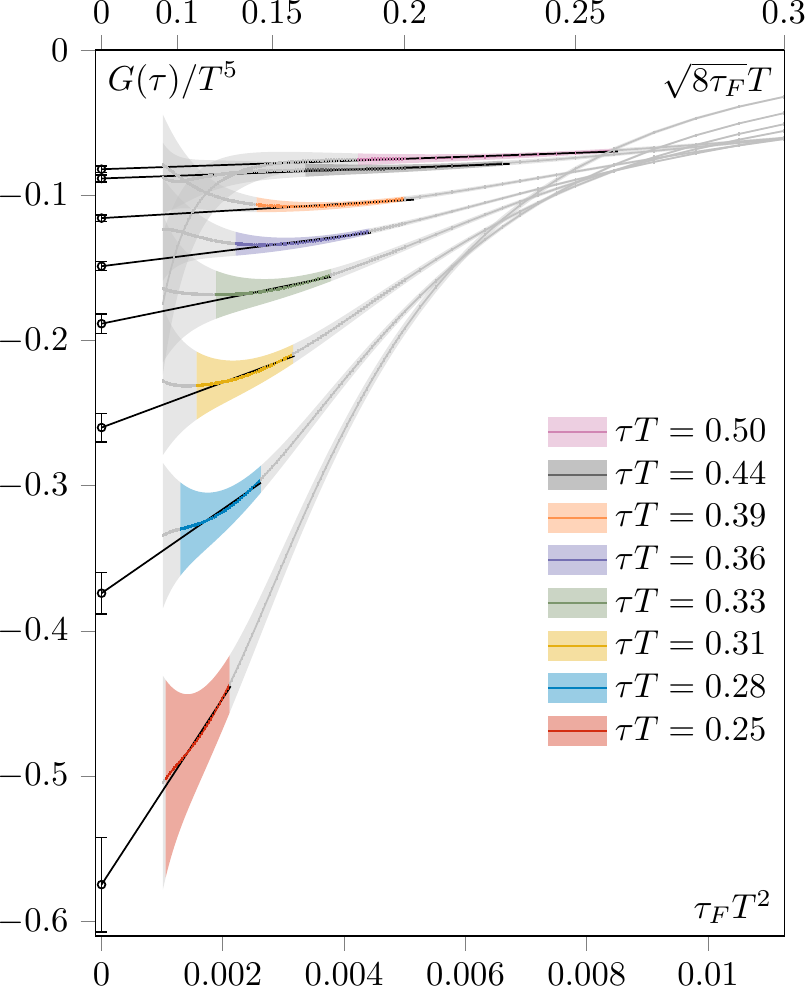}
    }
     \caption{Continuum-extrapolated topological charge density correlator
     as a function of flow time. Dark/light areas are the central value/one
     sigma error region.
     The colored parts are the ranges which are used to perform 
     the flow time extrapolations, while the black lines on 
     top indicate the linear extrapolations.
    The gray areas are not considered in the flow time extrapolation.
    }  \label{fig:qq_inverse_cont_corr}
\end{figure}
\subsection{Flow-time-to-zero extrapolation}

In order to perform an extrapolation to zero flow time, we first need to decide on the flow time range for the extrapolation.
In Ref.~\cite{Eller:2018yje} the authors propose a criterion for the upper limit of the flow time at given separation; it is determined by allowing the leading-order term of
$\langle F^a_{\mu\nu}\tilde{F}^a_{\mu\nu}(\tau)F^b_{\alpha\beta}\tilde{F}^b_{\alpha\beta}(0)\rangle$ to differ from its nonflowed counterpart by at most $1\%$.
For our nonperturbative data this constraint appears to be overly strict: the correlator's $\tf$ dependence is still moderate even for somewhat larger $\tf$ values, which can therefore be used in the small flow-time extrapolation. We loosen the criterion to a $20\%$ deviation, which leads to a maximum flow time within which a linear extrapolation is applicable, namely
$\sqrt{8\tau_{F}^{\mathrm{max}}}T=0.5220 \tau T$.
Because very small $\tf$ values are noisy, we then fit in the
range
\begin{align}\label{eq:flowlimits}
\tau_F\in[0.5\tau_{F}^{\mathrm{max}} , \tau_{F}^{\mathrm{max}} ].
\end{align}

A general analysis of continuum operators on gradient-flowed
field configurations shows that, in terms of unflowed operators,
they can be expanded as an operator product expansion,
and correspond to the desired operator, 
possibly with a renormalization factor,
plus a series of high-dimension operators with compensating positive
powers of $\tf$ as determined by operator dimension;
see for instance Refs \cite{Suzuki:2021tlr,Kitazawa:2017qab}.
In our case we know that the topological charge does not renormalize
(as we easily verify by seeing that it integrates to an integer)
and that the high-dimension operators must vanish on space integration,
implying that they are of form, e.g., $\tf D^2 q$.
Such a contaminating high-dimension operator does not affect
the determined total topology $\int q d^4 x$, but it does
affect the correlation functions, leading to corrections
of order $\tf / \tau^2$ based on dimensional reasoning.
Therefore, an appropriate Ansatz for our correlation function
at small $\tf$, incorporating the lowest-order corrections, is
\begin{equation}
\label{flow-extrap}
\frac{G_{\tau}(\tau_F)}{T^5}=c\cdot\tau_FT^2+d,
\end{equation}
where $d$ is the correlator at zero flow time, again normalized by $T^5$. The extrapolation procedure is illustrated in \autoref{fig:qq_inverse_cont_corr}. Each curve represents the continuum-extrapolated $G_{\tf}(\tau)$ at a fixed separation $\tau$, as a function of $\tf$.
The areas which are used for the flow-time-to-zero extrapolation according to \cref{eq:flowlimits} are colored.
The gray areas are not used in the extrapolation. 
The straight lines in black indicate the linear flow-time-to-zero extrapolation, while the discrete points at $\tf T^2=0$ are the final extrapolated values at $\tf=0$.  We see that the
extrapolated value and the value at the largest-used $\tf$
differ by at most 20\%, indicating that a small-$\tf$ expansion
should still be valid and that higher-order $\tf^2$ terms
should only give a few percent corrections and are not (yet)
needed.

\begin{figure}[t!]
\centerline{\includegraphics[width=0.46\textwidth]{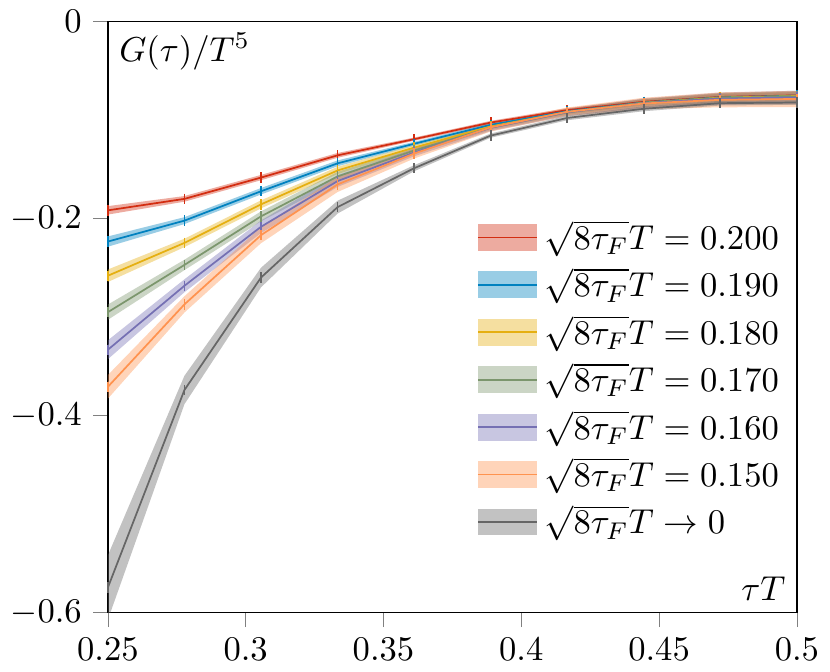}
    }
    \caption{Continuum extrapolated topological charge density correlator as well as the final 
    flow-time-extrapolated correlator.
    }  \label{fig:qq_cont_flow_extr}
\end{figure}

The topological charge density $q(x)$ is odd under time reflections. Due to arguments based on reflection positivity~\cite{Seiler:2001je,Vicari:1999xx},
the correlation function has to be negative for all nonzero separations $\tau T \neq 0$ in the continuum~\cite{Horvath:2005cv,Seiler:2001je}.  \autoref{fig:qq_cont_flow_extr} shows our final continuum and flow time extrapolated correlator. We indeed observe that the correlator is negative in the range we are analyzing, that is $\tau T > 0$.

\begin{figure}[t!] 
\centerline{\includegraphics[width=0.5\textwidth]{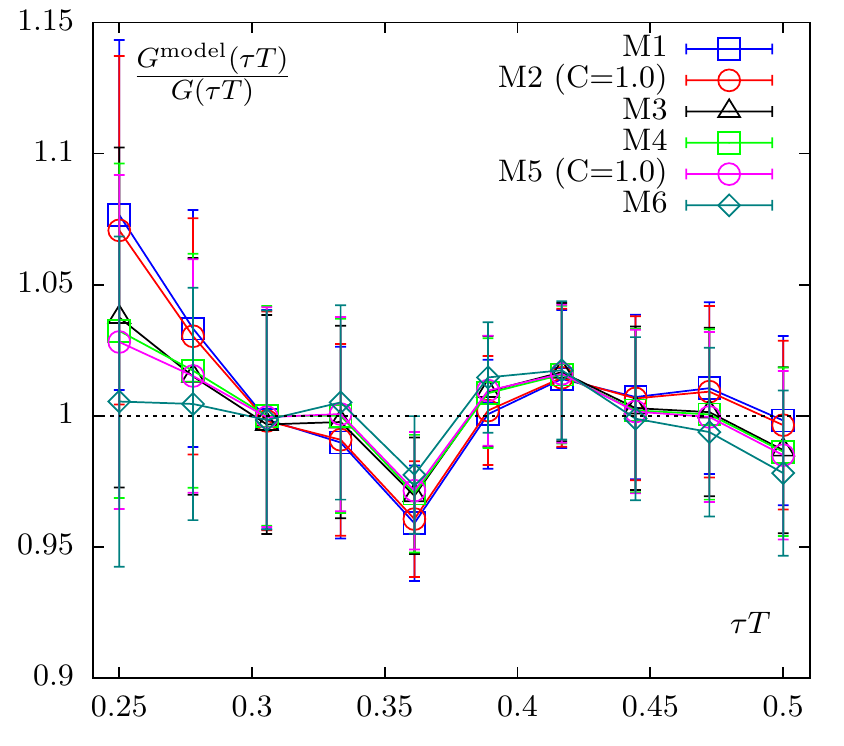}
}
\caption{The ratio of fit correlators to lattice data for different models. 
}
\label{fit_corrs}
\end{figure}

\section{Sphaleron rate}
\label{spha}

The sphaleron rate is determined by the small-frequency limiting behavior
of the spectral function as shown in
\cref{spha_formula},
but the Euclidean function receives large contributions from the
high-frequency tail of the spectral function, particularly at smaller
temporal separations, as shown in \cref{gqq}.
Therefore we need to incorporate as much information
as we can about this large-frequency region.  Fortunately, the behavior
at large frequencies should be more perturbative than the small-frequency part, and
the spectral function has been computed both at leading order (LO)
and at next-to-leading order (NLO) in the coupling in Ref.~\cite{Laine:2011xm}:
\begin{align}
\label{rhoLO}
\rho^{\mathrm{LO}}(\omega)= & 
 \frac{d_A \omega^4g^4}{2^{12}\pi^5}
 \coth\Big{(}\frac{\omega}{4T}\Big{)},\\
\rho^{\mathrm{NLO}}(\omega)= & \rho^{\mathrm{LO}} (\omega) +
 \frac{d_A \omega^4}{2^{12} \pi^5}
 \coth\Big{(}\frac{\omega}{4T}\Big{)} \frac{g^6(\bar{\mu})\Nc}{(4\pi)^2}  \nonumber \\
   &\times \biggl[
       \frac{22}{3} \ln\frac{\bar{\mu}^2}{\omega^2} + \frac{97}{3} 
     + 8\, \phi^{ }_T(\omega) 
    \biggr] .
 \label{rhofinal_chi}
\end{align}
Note that our definition of the spectral function differs from
that in Ref.~\cite{Laine:2011xm} by a relative minus sign.
Here $\dA = N_c^2-1=8$ is the dimension of the adjoint representation, which counts the number of gluon states.
At leading order one does not obtain a prescription on determining the value of the running coupling. We therefore make an educated guess of the renormalization point choosing the value from the 1-loop order ``EQCD'' setup yielding (Eq.(5.26) in \cite{Laine:2011xm})
\begin{align}
 \ln\left( \bar{\mu}^{\mathrm{opt}(T)} \right) \equiv \ln\left( 4\pi T\right) -\gamma_{\mathrm{E}} -\frac{1}{22} \,.
 \label{mu opt from T}
\end{align}
Using this relation the coupling is fixed to the value $g^2\left( \bar{\mu}^{\mathrm{opt}(T)}\right)=2.2346$ at $T=1.5T_c$, where we use an updated relation $T_c=1.24\Lambda_{\overline{\mathrm{MS}}}$~\cite{Francis:2015lha}.
We introduce an overall scaling coefficient $B$ in the LO models to compensate for the possibly bad choice of value of the coupling constant. As determined in \cite{Laine:2011xm}, at NLO, see \cref{rhofinal_chi}, the optimization of the scale $\bar{\mu}$ and the running of the strong coupling constant with $\omega$ is possible in the regime $\omega \gg \pi T$. In this regime the function $\phi_T(\omega)$, defined in Eq.(4.4) of \cite{Laine:2011xm}, is small which allows to define the optimized renormalization point as a function of $\omega$ as (Eq.(5.25) in \cite{Laine:2011xm})
\begin{align}
\ln\left( \bar{\mu}^{\mathrm{opt}(\omega)} \right) \equiv \ln\left(  \omega \right)  -\frac{97}{44} \,.
\label{mu opt from omega}
\end{align}
For values of $\omega$ outside this regime one again falls back to the renormalization point given by \cref{mu opt from T}.
Following the prescription given in \cite{Laine:2011xm}, one uses the larger value of \cref{mu opt from T} and \cref{mu opt from omega} for given $\omega$. The switch between the two formulations happens at $\omega/T=19.456 \pi$.

However, we also introduce the scaling parameter $B$ in the NLO models in order to compensate for higher order corrections to the value of the  renormalization point as well as other uncertainties in renormalization. Because the perturbative series is not rapidly converging
at this temperature, we fit the lattice data using both the LO
and the NLO spectral function, considering
the difference as an estimate of the uncertainties which arise due to our
incomplete knowledge of the spectral function's
high frequency functional form.

First, we consider a fit to just the leading-order spectral
function.  The fit is
poor, with $\chi^2/\mathrm{d.o.f} =68.6$. Replacing the
leading-order spectral function with the NLO expression from
\cref{rhofinal_chi}, even allowing for an additional
multiplicative overall rescaling $B$, also gives a poor fit,
with $\chi^2/\mathrm{d.o.f.} = 33.2$. Therefore it is necessary
(and theoretically justified)
to incorporate an additional structure, representing a low
frequency contribution to the spectral function.
Due to a lack of theoretical guidance, we will consider
three possibilities.  The first is a simple $\delta$-peak
in $\rho/\omega$ whose overall coefficient $A/T^4$ is treated
as another fit parameter.  This model is
theoretically motivated by the appearance
of a sharp feature in perturbative calculations
\cite{Arnold:1996dy}. We call this Model M1 when combined
with the LO given in \cref{rhoLO} times the scaling parameter $B$ and M4 when combined with 
$B$ times the NLO spectral function from \cref{rhofinal_chi}. Because the actual coupling is rather large, the assumption
of a sharp peak may not be reliable, so we also consider
a model in which the peak is broadened into a Breit-Wigner
distribution, e.g., $\rho_{\mathrm{peak}}/(\omega T^3)= (A/T^4) CT^2/(C^2T^2+\omega^2)$.
We treat $A/T^4$ as a fitting parameter and consider a few
distinct $C$ values, varying from a rather narrow to a quite
wide structure.  We call this model M2 when combined with
\cref{rhoLO} and M5 when combined with \cref{rhofinal_chi}.
Finally we consider the large-width
limit of this form, $\rho_{\mathrm{peak}}/(\omega T^3)= A/T^4$,
which is model M3 or M6.  In summary, our models are:
\begin{figure*}[t!] 
\centerline{\includegraphics[width=0.5\textwidth]{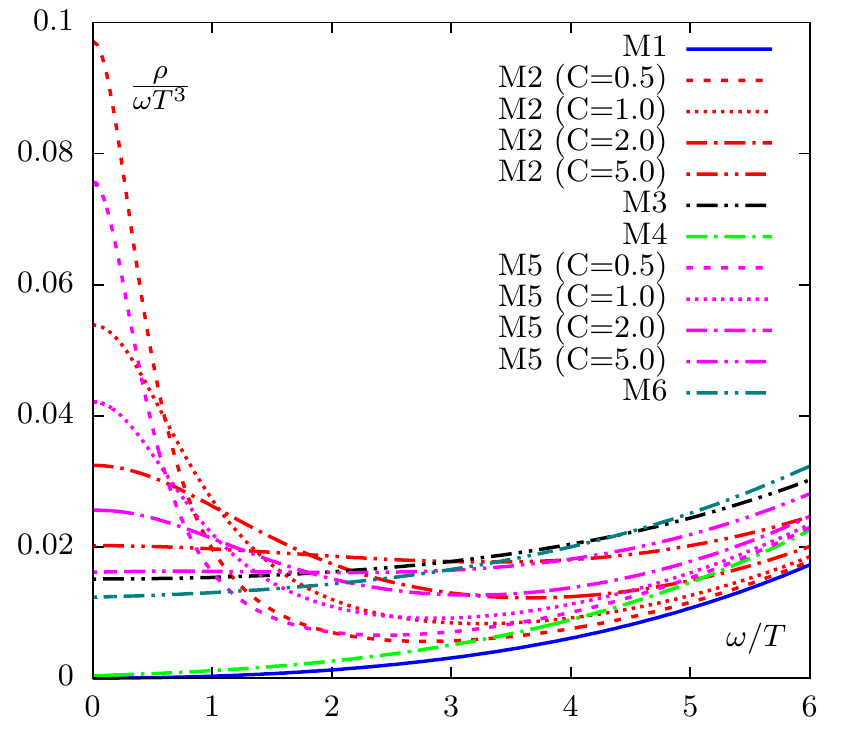}
\includegraphics[width=0.5\textwidth]{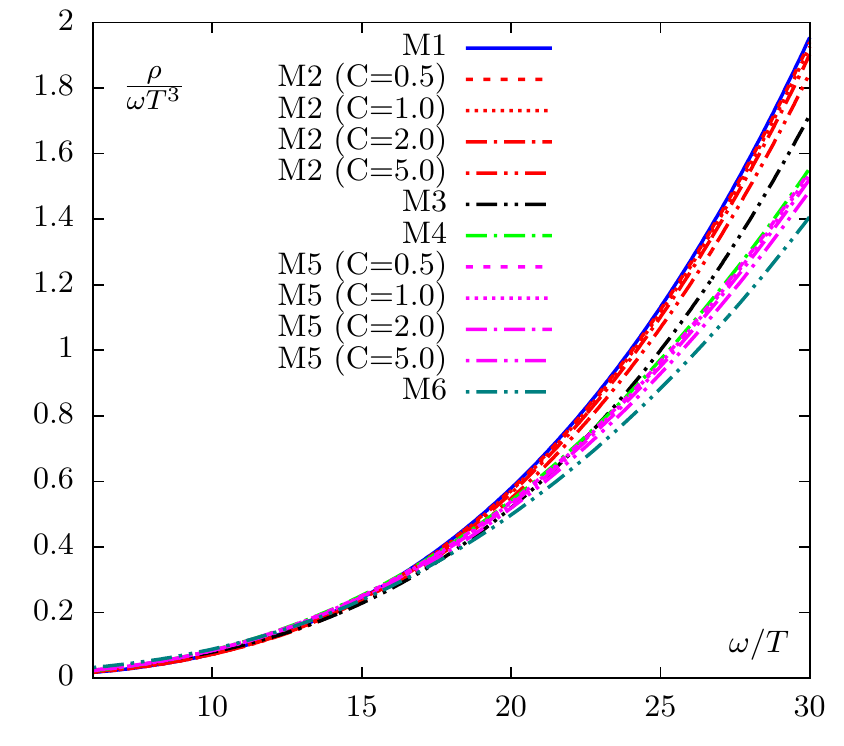}}
\caption{Left: small frequency parts of the fit spectral functions from different models. For M1 and M4 there is a $\delta$-peak at zero frequency, which is invisible in the plot. Right: same as the left panel but in larger $\omega/T$ range.
}
\label{fit_spf}
\end{figure*}

\begin{align*}
\label{models}
\mathrm{M1}: \frac{\rho(\omega)}
{\omega T^3}=&\frac{A}{T^4}\delta\left(\frac{\omega}{T} \right)
+B\frac{\rho^{\mathrm{LO}}(\omega)}{\omega T^3} \nonumber\\
\mathrm{M2}: \frac{\rho(\omega)}{\omega T^3}=
&\frac{A}{T^4}\frac{CT^2}{C^2T^2+\omega^2}+B\frac{\rho^{\mathrm{LO}}(\omega)}{\omega T^3}\nonumber\\
\mathrm{M3}: \frac{\rho(\omega)}{\omega T^3}=
&\frac{A}{T^4}+B\frac{\rho^{\mathrm{LO}}(\omega)}{\omega T^3}\nonumber\\
\end{align*}
\begin{align}
\mathrm{M4}: \frac{\rho(\omega)}{\omega T^3}=
&\frac{A}{T^4}\delta \left( \frac{\omega}{T} \right)+B\frac{\rho^{\mathrm{NLO}}(\omega)}{\omega T^3}\nonumber\\ 
\mathrm{M5}: \frac{\rho(\omega)}{\omega T^3}=
&\frac{A}{T^4}\frac{CT^2}{C^2 T^2+\omega^2}+B\frac{\rho^{\mathrm{NLO}}(\omega)}{\omega T^3}\nonumber\\
\mathrm{M6}: \frac{\rho(\omega)}{\omega T^3}=
&\frac{A}{T^4}+B\frac{\rho^{\mathrm{NLO}}(\omega)}{\omega T^3}.
\end{align}
The sphaleron rate is $2T^4$ times the value of the right-hand side at $\omega=0$.
For M2 and M5 the sphaleron rate can be calculated as $\Gamma_{\mathrm{sphal}}/T^4=2A/CT^4$ while for M3 and M6 it is $\Gamma_{\mathrm{sphal}}/T^4=2A/T^4$. For M1 and M4, the $\delta$-peak leads to an infinite sphaleron rate.

The fit results are summarized in \autoref{tabel2}.
In each case, our data rather tightly constraints
the fit parameters, and the quality of the fit
($\chi^2/\mathrm{d.o.f.}$) is fairly good.  The fits based
on the NLO high-frequency spectral function are slightly better than those using the LO spectral function, but
no model for the low-frequency behavior is clearly preferred.
We also show the ratio of fit correlators to the lattice data, and the resulting spectral functions in \autoref{fit_corrs} and \autoref{fit_spf}.

We use the Levenberg-Marquardt algorithm for $\chi^2$-fitting and stop the iteration when a relative tolerance of $10^{-7}$ is reached for the fit parameters or the $\chi^2/\mathrm{d.o.f}$.
Only data points at or beyond $\tau T = 0.25$ are used in the fit, since the transport information is mainly encoded in the correlator at large separations and the $\tf$-window over which we can perform the flow-time-to-zero extrapolation closes up as we go to shorter and shorter separations.

\begin{figure}[t!] 
\centerline{\includegraphics[width=0.5\textwidth]{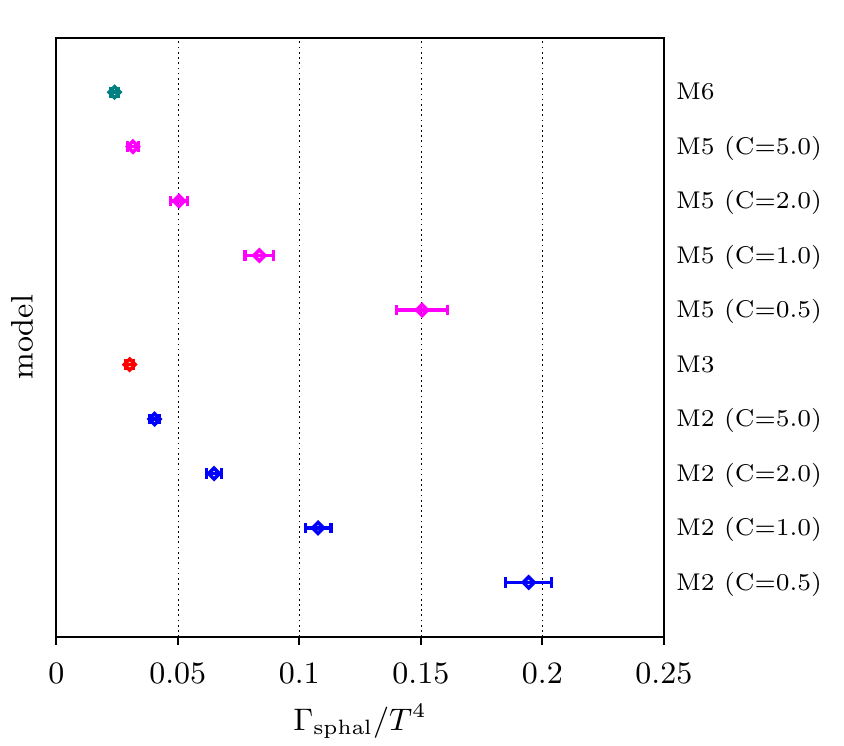}}
\caption{The sphaleron rate from different models at $1.5\,T_c$.}
\label{spha_model}
\end{figure}

Now we calculate the sphaleron rate in different models according to \cref{spha_formula} and summarize them in \autoref{spha_model}. 
From our analysis we can see  that a $\delta$-like transport peak
and a linear-in-frequency transport peak are just special cases
of a Breit-Wigner transport peak with zero and infinite width.
Within the set of fit models we have considered, we find
that the sphaleron rate varies within the range
\begin{equation}
    \begin{split}
        &\Gamma_{\mathrm{spha}}/T^4 \geq 0.030(2),\ \ \textrm{based on M1-M3},\\
        &\Gamma_{\mathrm{spha}}/T^4 \geq 0.024(2),\ \ \textrm{based on M4-M6}.
    \end{split}
\end{equation}
The first line is based on the LO high-frequency spectral function and the second line is based on the NLO spectral function.

Note that our results indicate that a spectral function with only a high-frequency part does not provide a good fit;
there \textsl{must} be a low-frequency structure in addition.
The functional forms we have considered contain two extremes
for such a form; an infinitely sharp peak (M1/M4) and an
infinitely broad peak (M3/M6).  Therefore we consider them to
span the range of likely functional forms for a ``peak,'' and
we propose that the lower value found can be viewed as a lower
limit on the sphaleron rate.  We are aware that this claim
depends somewhat on our choice of fitting functions considered,
but again we have considered a broad range with peak structures
ranging from sharp to perfectly wide, so we consider our
bound to be reasonable.

\begin{table*}[t!]
\centering
\begingroup
\setlength{\tabcolsep}{5pt}
\renewcommand{\arraystretch}{1.3}
\begin{tabular}{cccccc}
\hline  \hline
        Ansatz & $A/T^4\times 10$ & $B$ & $C$ & $\chi^2/$d.o.f & $\Gamma_{\mathrm{sphal}}/T^4\times 10$
            \tabularnewline
            \hline
$\mathrm{M1}$ & 0.68(4) & 2.27(7) & — & 1.86 & $\infty$ \tabularnewline
$\mathrm{M2}$ & 0.49(3) & 2.25(7) & 0.5 & 2.07 & 1.94(10) \tabularnewline
$\mathrm{M2}$ & 0.54(3) & 2.24(7) & 1.0 & 2.02 & 1.08(6) \tabularnewline
$\mathrm{M2}$ & 0.65(4) & 2.21(7) & 2.0 & 1.93 & 0.65(4) \tabularnewline
$\mathrm{M2}$ & 1.01(5) & 2.14(7) & 5.0 & 1.76 & 0.40(2) \tabularnewline
$\mathrm{M3}$ & 0.15(1) & 1.98(8) & — & 1.36 & 0.30(2) \tabularnewline
$\mathrm{M4}$ & 0.53(4) & 1.25(4) & —  & 1.35 & $\infty$ \tabularnewline
$\mathrm{M5}$ & 0.38(3) & 1.25(4) & 0.5 & 1.53 & 1.50(11) \tabularnewline
$\mathrm{M5}$ & 0.42(3) & 1.24(4) & 1.0 & 1.51 & 0.84(6) \tabularnewline
$\mathrm{M5}$ & 0.50(4) & 1.23(4) & 2.0 & 1.48 & 0.50(4) \tabularnewline
$\mathrm{M5}$ & 0.79(6) & 1.20(4) & 5.0 & 1.47 & 0.32(3) \tabularnewline
$\mathrm{M6}$ & 0.12(1) & 1.12(5) & — & 1.29 & 0.24(2)
\tabularnewline
\hline
\hline
\end{tabular}
\endgroup
\caption{Fitted parameters and $\chi^2/\mathrm{d.o.f}$ for different ansätze. The numbers in the parentheses are statistical uncertainties from a bootstrap analysis. The sphaleron rate $\Gamma_{\mathrm{sphal}}/T^4$ is calculated from the fitted parameters $A/T^4$ and $C$. The meaning of each fit parameter and how the sphaleron rate is determined in each ansatz can be found in \autoref{spha}. In ansatz M1-M3 the LO spectral function has been used at large frequencies while for M4-M6 the NLO spectral function has been used.}
\label{tabel2}
\end{table*}

\section{Conclusion}
\label{sec:conclusion}

We have calculated the topological charge correlation functions on 5 different large and fine isotropic lattices in pure-glue QCD
at a temperature $T = 1.5 T_c$.
To improve the signal we used the gradient flow method.
Within the framework of the gradient flow we developed a methodology to perform reliable double-extrapolations for the topological charge correlators.
The correlators extrapolated to the $a\rightarrow 0$ and $\tau_F\rightarrow 0$ limit are then used in the spectral reconstruction,
where we use perturbatively inspired models considering uncertainties from different sources.
Using either the LO or the NLO spectral function \textsl{alone}
leads to a very poor fit; the data demand the addition of a
low-frequency structure.  By considering a range of such structures,
the lowest value we can obtain for the sphaleron rate, for the
case that the added structure is completely flat, is
$\Gamma_{\mathrm{spha}}/T^4 \geq 0.024$ at $1.5T_c$.
Low-frequency structures containing an actual peak give higher
values for $\Gamma_{\mathrm{spha}}$.
If only the LO spectral function is used in the fit, then a stronger constraint,
$\Gamma_{\mathrm{spha}}/T^4 \geq 0.030$ is obtained.

In our opinion there are a few take-home messages from our work.
First, the use of gradient flow is essential to ensure that the topological
charge correctly captures topological information.  However, only a limited
flow time range is useful for the flow-time-to-zero extrapolation.  At small separations and coarse lattices these constraints cannot both be maintained, and we are driven to very fine lattice spacings and can only use the larger available separations. Second, although we were able to extract continuum and flow-time-to-zero extrapolated data with a few percent statistical error bars,
the lack of clear theoretical guidance in fitting the spectral function leads to rather severe theory errors in the resulting sphaleron rate.
We can place a lower bound on the sphaleron rate, but not a useful upper bound.  
It would be useful to improve our theoretical understanding of the expected form for the topological charge-density spectral function, particularly in the low-frequency region;
such improvements might allow a reasonably good determination of the sphaleron rate.

All data from our calculations, presented in the figures of this paper, can be found in \cite{datapublication}.

\begin{acknowledgments}

We thank Mikko Laine for a very helpful communication.
All authors acknowledge support by the Deutsche For\-schungs\-ge\-mein\-schaft
(DFG, German Research Foundation) through the CRC-TR 211 'Strong-interaction matter under extreme conditions'– project number 315477589 – TRR 211.
The computations in this work were performed on the GPU cluster at Bielefeld University.

\end{acknowledgments}

\bibliography{Bibliography}

\begin{thebibliography}{51}%
\makeatletter
\providecommand \@ifxundefined [1]{%
 \@ifx{#1\undefined}
}%
\providecommand \@ifnum [1]{%
 \ifnum #1\expandafter \@firstoftwo
 \else \expandafter \@secondoftwo
 \fi
}%
\providecommand \@ifx [1]{%
 \ifx #1\expandafter \@firstoftwo
 \else \expandafter \@secondoftwo
 \fi
}%
\providecommand \natexlab [1]{#1}%
\providecommand \enquote  [1]{``#1''}%
\providecommand \bibnamefont  [1]{#1}%
\providecommand \bibfnamefont [1]{#1}%
\providecommand \citenamefont [1]{#1}%
\providecommand \href@noop [0]{\@secondoftwo}%
\providecommand \href [0]{\begingroup \@sanitize@url \@href}%
\providecommand \@href[1]{\@@startlink{#1}\@@href}%
\providecommand \@@href[1]{\endgroup#1\@@endlink}%
\providecommand \@sanitize@url [0]{\catcode `\\12\catcode `\$12\catcode
  `\&12\catcode `\#12\catcode `\^12\catcode `\_12\catcode `\%12\relax}%
\providecommand \@@startlink[1]{}%
\providecommand \@@endlink[0]{}%
\providecommand \url  [0]{\begingroup\@sanitize@url \@url }%
\providecommand \@url [1]{\endgroup\@href {#1}{\urlprefix }}%
\providecommand \urlprefix  [0]{URL }%
\providecommand \Eprint [0]{\href }%
\providecommand \doibase [0]{http://dx.doi.org/}%
\providecommand \selectlanguage [0]{\@gobble}%
\providecommand \bibinfo  [0]{\@secondoftwo}%
\providecommand \bibfield  [0]{\@secondoftwo}%
\providecommand \translation [1]{[#1]}%
\providecommand \BibitemOpen [0]{}%
\providecommand \bibitemStop [0]{}%
\providecommand \bibitemNoStop [0]{.\EOS\space}%
\providecommand \EOS [0]{\spacefactor3000\relax}%
\providecommand \BibitemShut  [1]{\csname bibitem#1\endcsname}%
\let\auto@bib@innerbib\@empty
\bibitem [{\citenamefont {McLerran}\ \emph {et~al.}(1991)\citenamefont
  {McLerran}, \citenamefont {Mottola},\ and\ \citenamefont
  {Shaposhnikov}}]{McLerran:1990de}%
  \BibitemOpen
  \bibfield  {author} {\bibinfo {author} {\bibfnamefont {L.~D.}\ \bibnamefont
  {McLerran}}, \bibinfo {author} {\bibfnamefont {E.}~\bibnamefont {Mottola}}, \
  and\ \bibinfo {author} {\bibfnamefont {M.~E.}\ \bibnamefont {Shaposhnikov}},\
  }\href {\doibase 10.1103/PhysRevD.43.2027} {\bibfield  {journal} {\bibinfo
  {journal} {Phys. Rev. D}\ }\textbf {\bibinfo {volume} {43}},\ \bibinfo
  {pages} {2027} (\bibinfo {year} {1991})}\BibitemShut {NoStop}%
\bibitem [{\citenamefont {Giudice}\ and\ \citenamefont
  {Shaposhnikov}(1994)}]{Giudice:1993bb}%
  \BibitemOpen
  \bibfield  {author} {\bibinfo {author} {\bibfnamefont {G.}~\bibnamefont
  {Giudice}}\ and\ \bibinfo {author} {\bibfnamefont {M.~E.}\ \bibnamefont
  {Shaposhnikov}},\ }\href {\doibase 10.1016/0370-2693(94)91202-5} {\bibfield
  {journal} {\bibinfo  {journal} {Phys. Lett. B}\ }\textbf {\bibinfo {volume}
  {326}},\ \bibinfo {pages} {118} (\bibinfo {year} {1994})}\BibitemShut
  {NoStop}%
\bibitem [{\citenamefont {Kharzeev}\ \emph {et~al.}(2008)\citenamefont
  {Kharzeev}, \citenamefont {McLerran},\ and\ \citenamefont
  {Warringa}}]{Kharzeev:2007jp}%
  \BibitemOpen
  \bibfield  {author} {\bibinfo {author} {\bibfnamefont {D.~E.}\ \bibnamefont
  {Kharzeev}}, \bibinfo {author} {\bibfnamefont {L.~D.}\ \bibnamefont
  {McLerran}}, \ and\ \bibinfo {author} {\bibfnamefont {H.~J.}\ \bibnamefont
  {Warringa}},\ }\href {\doibase 10.1016/j.nuclphysa.2008.02.298} {\bibfield
  {journal} {\bibinfo  {journal} {Nucl. Phys. A}\ }\textbf {\bibinfo {volume}
  {803}},\ \bibinfo {pages} {227} (\bibinfo {year} {2008})}\BibitemShut
  {NoStop}%
\bibitem [{\citenamefont {Fukushima}\ \emph {et~al.}(2008)\citenamefont
  {Fukushima}, \citenamefont {Kharzeev},\ and\ \citenamefont
  {Warringa}}]{Fukushima:2008xe}%
  \BibitemOpen
  \bibfield  {author} {\bibinfo {author} {\bibfnamefont {K.}~\bibnamefont
  {Fukushima}}, \bibinfo {author} {\bibfnamefont {D.~E.}\ \bibnamefont
  {Kharzeev}}, \ and\ \bibinfo {author} {\bibfnamefont {H.~J.}\ \bibnamefont
  {Warringa}},\ }\href {\doibase 10.1103/PhysRevD.78.074033} {\bibfield
  {journal} {\bibinfo  {journal} {Phys. Rev. D}\ }\textbf {\bibinfo {volume}
  {78}},\ \bibinfo {pages} {074033} (\bibinfo {year} {2008})}\BibitemShut
  {NoStop}%
\bibitem [{\citenamefont {Moore}\ and\ \citenamefont
  {Tassler}(2011)}]{Moore:2010jd}%
  \BibitemOpen
  \bibfield  {author} {\bibinfo {author} {\bibfnamefont {G.~D.}\ \bibnamefont
  {Moore}}\ and\ \bibinfo {author} {\bibfnamefont {M.}~\bibnamefont
  {Tassler}},\ }\href {\doibase 10.1007/JHEP02(2011)105} {\bibfield  {journal}
  {\bibinfo  {journal} {JHEP}\ }\textbf {\bibinfo {volume} {02}},\ \bibinfo
  {pages} {105} (\bibinfo {year} {2011})}\BibitemShut {NoStop}%
\bibitem [{\citenamefont {'t~Hooft}(1976)}]{tHooft:1976snw}%
  \BibitemOpen
  \bibfield  {author} {\bibinfo {author} {\bibfnamefont {G.}~\bibnamefont
  {'t~Hooft}},\ }\href {\doibase 10.1103/PhysRevD.14.3432} {\bibfield
  {journal} {\bibinfo  {journal} {Phys. Rev. D}\ }\textbf {\bibinfo {volume}
  {14}},\ \bibinfo {pages} {3432} (\bibinfo {year} {1976})},\ \bibinfo {note}
  {[Erratum: Phys.Rev.D 18, 2199 (1978)]}\BibitemShut {NoStop}%
\bibitem [{\citenamefont {Atiyah}\ and\ \citenamefont
  {Singer}(1968)}]{Atiyah:1968mp}%
  \BibitemOpen
  \bibfield  {author} {\bibinfo {author} {\bibfnamefont {M.}~\bibnamefont
  {Atiyah}}\ and\ \bibinfo {author} {\bibfnamefont {I.}~\bibnamefont
  {Singer}},\ }\href {\doibase 10.2307/1970715} {\bibfield  {journal} {\bibinfo
   {journal} {Annals Math.}\ }\textbf {\bibinfo {volume} {87}},\ \bibinfo
  {pages} {484} (\bibinfo {year} {1968})}\BibitemShut {NoStop}%
\bibitem [{\citenamefont {Kharzeev}\ \emph {et~al.}(2020)\citenamefont
  {Kharzeev}, \citenamefont {Shuryak},\ and\ \citenamefont
  {Zahed}}]{Kharzeev:2019rsy}%
  \BibitemOpen
  \bibfield  {author} {\bibinfo {author} {\bibfnamefont {D.}~\bibnamefont
  {Kharzeev}}, \bibinfo {author} {\bibfnamefont {E.}~\bibnamefont {Shuryak}}, \
  and\ \bibinfo {author} {\bibfnamefont {I.}~\bibnamefont {Zahed}},\ }\href
  {\doibase 10.1103/PhysRevD.102.073003} {\bibfield  {journal} {\bibinfo
  {journal} {Phys. Rev. D}\ }\textbf {\bibinfo {volume} {102}},\ \bibinfo
  {pages} {073003} (\bibinfo {year} {2020})}\BibitemShut {NoStop}%
\bibitem [{\citenamefont {Voloshin}(2010)}]{Voloshin:2010ut}%
  \BibitemOpen
  \bibfield  {author} {\bibinfo {author} {\bibfnamefont {S.~A.}\ \bibnamefont
  {Voloshin}},\ }\href {\doibase 10.1103/PhysRevLett.105.172301} {\bibfield
  {journal} {\bibinfo  {journal} {Phys. Rev. Lett.}\ }\textbf {\bibinfo
  {volume} {105}},\ \bibinfo {pages} {172301} (\bibinfo {year}
  {2010})}\BibitemShut {NoStop}%
\bibitem [{\citenamefont {Bodeker}(1998)}]{Bodeker:1998hm}%
  \BibitemOpen
  \bibfield  {author} {\bibinfo {author} {\bibfnamefont {D.}~\bibnamefont
  {Bodeker}},\ }\href {\doibase 10.1016/S0370-2693(98)00279-2} {\bibfield
  {journal} {\bibinfo  {journal} {Phys. Lett. B}\ }\textbf {\bibinfo {volume}
  {426}},\ \bibinfo {pages} {351} (\bibinfo {year} {1998})}\BibitemShut
  {NoStop}%
\bibitem [{\citenamefont {Bodeker}(2000)}]{Bodeker:1999ud}%
  \BibitemOpen
  \bibfield  {author} {\bibinfo {author} {\bibfnamefont {D.}~\bibnamefont
  {Bodeker}},\ }\href {\doibase 10.1016/S0550-3213(99)00582-9} {\bibfield
  {journal} {\bibinfo  {journal} {Nucl. Phys. B}\ }\textbf {\bibinfo {volume}
  {566}},\ \bibinfo {pages} {402} (\bibinfo {year} {2000})}\BibitemShut
  {NoStop}%
\bibitem [{\citenamefont {Bodeker}(1999)}]{Bodeker:1999ey}%
  \BibitemOpen
  \bibfield  {author} {\bibinfo {author} {\bibfnamefont {D.}~\bibnamefont
  {Bodeker}},\ }\href {\doibase 10.1016/S0550-3213(99)00435-6} {\bibfield
  {journal} {\bibinfo  {journal} {Nucl. Phys. B}\ }\textbf {\bibinfo {volume}
  {559}},\ \bibinfo {pages} {502} (\bibinfo {year} {1999})}\BibitemShut
  {NoStop}%
\bibitem [{\citenamefont {Arnold}\ and\ \citenamefont
  {Yaffe}(2000)}]{Arnold:1999uy}%
  \BibitemOpen
  \bibfield  {author} {\bibinfo {author} {\bibfnamefont {P.~B.}\ \bibnamefont
  {Arnold}}\ and\ \bibinfo {author} {\bibfnamefont {L.~G.}\ \bibnamefont
  {Yaffe}},\ }\href {\doibase 10.1103/PhysRevD.62.125014} {\bibfield  {journal}
  {\bibinfo  {journal} {Phys. Rev. D}\ }\textbf {\bibinfo {volume} {62}},\
  \bibinfo {pages} {125014} (\bibinfo {year} {2000})}\BibitemShut {NoStop}%
\bibitem [{\citenamefont {Moore}(2000)}]{Moore:1998zk}%
  \BibitemOpen
  \bibfield  {author} {\bibinfo {author} {\bibfnamefont {G.~D.}\ \bibnamefont
  {Moore}},\ }\href {\doibase 10.1016/S0550-3213(99)00746-4} {\bibfield
  {journal} {\bibinfo  {journal} {Nucl. Phys. B}\ }\textbf {\bibinfo {volume}
  {568}},\ \bibinfo {pages} {367} (\bibinfo {year} {2000})}\BibitemShut
  {NoStop}%
\bibitem [{\citenamefont {Bodeker}\ \emph {et~al.}(2000)\citenamefont
  {Bodeker}, \citenamefont {Moore},\ and\ \citenamefont
  {Rummukainen}}]{Bodeker:1999gx}%
  \BibitemOpen
  \bibfield  {author} {\bibinfo {author} {\bibfnamefont {D.}~\bibnamefont
  {Bodeker}}, \bibinfo {author} {\bibfnamefont {G.~D.}\ \bibnamefont {Moore}},
  \ and\ \bibinfo {author} {\bibfnamefont {K.}~\bibnamefont {Rummukainen}},\
  }\href {\doibase 10.1103/PhysRevD.61.056003} {\bibfield  {journal} {\bibinfo
  {journal} {Phys. Rev. D}\ }\textbf {\bibinfo {volume} {61}},\ \bibinfo
  {pages} {056003} (\bibinfo {year} {2000})}\BibitemShut {NoStop}%
\bibitem [{\citenamefont {Kotov}(2018)}]{Kotov:2018vyl}%
  \BibitemOpen
  \bibfield  {author} {\bibinfo {author} {\bibfnamefont {A.}~\bibnamefont
  {Kotov}},\ }\href {\doibase 10.1134/S0021364018180078} {\bibfield  {journal}
  {\bibinfo  {journal} {JETP Lett.}\ }\textbf {\bibinfo {volume} {108}},\
  \bibinfo {pages} {352} (\bibinfo {year} {2018})}\BibitemShut {NoStop}%
\bibitem [{\citenamefont {Luscher}\ and\ \citenamefont
  {Weisz}(2011)}]{Luscher:2011bx}%
  \BibitemOpen
  \bibfield  {author} {\bibinfo {author} {\bibfnamefont {M.}~\bibnamefont
  {Luscher}}\ and\ \bibinfo {author} {\bibfnamefont {P.}~\bibnamefont
  {Weisz}},\ }\href {\doibase 10.1007/JHEP02(2011)051} {\bibfield  {journal}
  {\bibinfo  {journal} {JHEP}\ }\textbf {\bibinfo {volume} {02}},\ \bibinfo
  {pages} {051} (\bibinfo {year} {2011})}\BibitemShut {NoStop}%
\bibitem [{\citenamefont {Lüscher}(2010)}]{Luscher:2010iy}%
  \BibitemOpen
  \bibfield  {author} {\bibinfo {author} {\bibfnamefont {M.}~\bibnamefont
  {Lüscher}},\ }\href {\doibase 10.1007/JHEP08(2010)071} {\bibfield  {journal}
  {\bibinfo  {journal} {JHEP}\ }\textbf {\bibinfo {volume} {08}},\ \bibinfo
  {pages} {071} (\bibinfo {year} {2010})},\ \bibinfo {note} {[Erratum: JHEP 03,
  092 (2014)]}\BibitemShut {NoStop}%
\bibitem [{\citenamefont {Narayanan}\ and\ \citenamefont
  {Neuberger}(2006)}]{Narayanan:2006rf}%
  \BibitemOpen
  \bibfield  {author} {\bibinfo {author} {\bibfnamefont {R.}~\bibnamefont
  {Narayanan}}\ and\ \bibinfo {author} {\bibfnamefont {H.}~\bibnamefont
  {Neuberger}},\ }\href {\doibase 10.1088/1126-6708/2006/03/064} {\bibfield
  {journal} {\bibinfo  {journal} {JHEP}\ }\textbf {\bibinfo {volume} {03}},\
  \bibinfo {pages} {064} (\bibinfo {year} {2006})}\BibitemShut {NoStop}%
\bibitem [{\citenamefont {Berg}(1981)}]{BERG1981475}%
  \BibitemOpen
  \bibfield  {author} {\bibinfo {author} {\bibfnamefont {B.}~\bibnamefont
  {Berg}},\ }\href {\doibase https://doi.org/10.1016/0370-2693(81)90518-9}
  {\bibfield  {journal} {\bibinfo  {journal} {Physics Letters B}\ }\textbf
  {\bibinfo {volume} {104}},\ \bibinfo {pages} {475 } (\bibinfo {year}
  {1981})}\BibitemShut {NoStop}%
\bibitem [{\citenamefont {Iwasaki}\ and\ \citenamefont
  {Yoshiè}(1983)}]{IWASAKI1983159}%
  \BibitemOpen
  \bibfield  {author} {\bibinfo {author} {\bibfnamefont {Y.}~\bibnamefont
  {Iwasaki}}\ and\ \bibinfo {author} {\bibfnamefont {T.}~\bibnamefont
  {Yoshiè}},\ }\href {\doibase https://doi.org/10.1016/0370-2693(83)91111-5}
  {\bibfield  {journal} {\bibinfo  {journal} {Physics Letters B}\ }\textbf
  {\bibinfo {volume} {131}},\ \bibinfo {pages} {159 } (\bibinfo {year}
  {1983})}\BibitemShut {NoStop}%
\bibitem [{\citenamefont {Albanese}\ \emph {et~al.}(1987)\citenamefont
  {Albanese} \emph {et~al.}}]{Albanese:1987ds}%
  \BibitemOpen
  \bibfield  {author} {\bibinfo {author} {\bibfnamefont {M.}~\bibnamefont
  {Albanese}} \emph {et~al.} (\bibinfo {collaboration} {APE}),\ }\href
  {\doibase 10.1016/0370-2693(87)91160-9} {\bibfield  {journal} {\bibinfo
  {journal} {Phys. Lett. B}\ }\textbf {\bibinfo {volume} {192}},\ \bibinfo
  {pages} {163} (\bibinfo {year} {1987})}\BibitemShut {NoStop}%
\bibitem [{\citenamefont {Morningstar}\ and\ \citenamefont
  {Peardon}(2004)}]{Morningstar:2003gk}%
  \BibitemOpen
  \bibfield  {author} {\bibinfo {author} {\bibfnamefont {C.}~\bibnamefont
  {Morningstar}}\ and\ \bibinfo {author} {\bibfnamefont {M.~J.}\ \bibnamefont
  {Peardon}},\ }\href {\doibase 10.1103/PhysRevD.69.054501} {\bibfield
  {journal} {\bibinfo  {journal} {Phys. Rev. D}\ }\textbf {\bibinfo {volume}
  {69}},\ \bibinfo {pages} {054501} (\bibinfo {year} {2004})}\BibitemShut
  {NoStop}%
\bibitem [{\citenamefont {Hasenfratz}\ and\ \citenamefont
  {Knechtli}(2001)}]{Hasenfratz:2001hp}%
  \BibitemOpen
  \bibfield  {author} {\bibinfo {author} {\bibfnamefont {A.}~\bibnamefont
  {Hasenfratz}}\ and\ \bibinfo {author} {\bibfnamefont {F.}~\bibnamefont
  {Knechtli}},\ }\href {\doibase 10.1103/PhysRevD.64.034504} {\bibfield
  {journal} {\bibinfo  {journal} {Phys. Rev. D}\ }\textbf {\bibinfo {volume}
  {64}},\ \bibinfo {pages} {034504} (\bibinfo {year} {2001})}\BibitemShut
  {NoStop}%
\bibitem [{\citenamefont {Alexandrou}\ \emph {et~al.}(2015)\citenamefont
  {Alexandrou}, \citenamefont {Athenodorou},\ and\ \citenamefont
  {Jansen}}]{Alexandrou:2015yba}%
  \BibitemOpen
  \bibfield  {author} {\bibinfo {author} {\bibfnamefont {C.}~\bibnamefont
  {Alexandrou}}, \bibinfo {author} {\bibfnamefont {A.}~\bibnamefont
  {Athenodorou}}, \ and\ \bibinfo {author} {\bibfnamefont {K.}~\bibnamefont
  {Jansen}},\ }\href {\doibase 10.1103/PhysRevD.92.125014} {\bibfield
  {journal} {\bibinfo  {journal} {Phys. Rev. D}\ }\textbf {\bibinfo {volume}
  {92}},\ \bibinfo {pages} {125014} (\bibinfo {year} {2015})}\BibitemShut
  {NoStop}%
\bibitem [{\citenamefont {Alexandrou}\ \emph {et~al.}(2020)\citenamefont
  {Alexandrou}, \citenamefont {Athenodorou}, \citenamefont {Cichy},
  \citenamefont {Dromard}, \citenamefont {Garcia-Ramos}, \citenamefont
  {Jansen}, \citenamefont {Wenger},\ and\ \citenamefont
  {Zimmermann}}]{Alexandrou:2017hqw}%
  \BibitemOpen
  \bibfield  {author} {\bibinfo {author} {\bibfnamefont {C.}~\bibnamefont
  {Alexandrou}}, \bibinfo {author} {\bibfnamefont {A.}~\bibnamefont
  {Athenodorou}}, \bibinfo {author} {\bibfnamefont {K.}~\bibnamefont {Cichy}},
  \bibinfo {author} {\bibfnamefont {A.}~\bibnamefont {Dromard}}, \bibinfo
  {author} {\bibfnamefont {E.}~\bibnamefont {Garcia-Ramos}}, \bibinfo {author}
  {\bibfnamefont {K.}~\bibnamefont {Jansen}}, \bibinfo {author} {\bibfnamefont
  {U.}~\bibnamefont {Wenger}}, \ and\ \bibinfo {author} {\bibfnamefont
  {F.}~\bibnamefont {Zimmermann}},\ }\href {\doibase
  10.1140/epjc/s10052-020-7984-9} {\bibfield  {journal} {\bibinfo  {journal}
  {Eur. Phys. J. C}\ }\textbf {\bibinfo {volume} {80}},\ \bibinfo {pages} {424}
  (\bibinfo {year} {2020})}\BibitemShut {NoStop}%
\bibitem [{\citenamefont {Bonati}\ and\ \citenamefont
  {D'Elia}(2014)}]{Bonati:2014tqa}%
  \BibitemOpen
  \bibfield  {author} {\bibinfo {author} {\bibfnamefont {C.}~\bibnamefont
  {Bonati}}\ and\ \bibinfo {author} {\bibfnamefont {M.}~\bibnamefont
  {D'Elia}},\ }\href {\doibase 10.1103/PhysRevD.89.105005} {\bibfield
  {journal} {\bibinfo  {journal} {Phys. Rev. D}\ }\textbf {\bibinfo {volume}
  {89}},\ \bibinfo {pages} {105005} (\bibinfo {year} {2014})}\BibitemShut
  {NoStop}%
\bibitem [{\citenamefont {Kitazawa}\ \emph {et~al.}(2016)\citenamefont
  {Kitazawa}, \citenamefont {Iritani}, \citenamefont {Asakawa}, \citenamefont
  {Hatsuda},\ and\ \citenamefont {Suzuki}}]{Kitazawa:2016dsl}%
  \BibitemOpen
  \bibfield  {author} {\bibinfo {author} {\bibfnamefont {M.}~\bibnamefont
  {Kitazawa}}, \bibinfo {author} {\bibfnamefont {T.}~\bibnamefont {Iritani}},
  \bibinfo {author} {\bibfnamefont {M.}~\bibnamefont {Asakawa}}, \bibinfo
  {author} {\bibfnamefont {T.}~\bibnamefont {Hatsuda}}, \ and\ \bibinfo
  {author} {\bibfnamefont {H.}~\bibnamefont {Suzuki}},\ }\href {\doibase
  10.1103/PhysRevD.94.114512} {\bibfield  {journal} {\bibinfo  {journal} {Phys.
  Rev. D}\ }\textbf {\bibinfo {volume} {94}},\ \bibinfo {pages} {114512}
  (\bibinfo {year} {2016})}\BibitemShut {NoStop}%
\bibitem [{\citenamefont {Kitazawa}\ \emph {et~al.}(2017)\citenamefont
  {Kitazawa}, \citenamefont {Iritani}, \citenamefont {Asakawa},\ and\
  \citenamefont {Hatsuda}}]{Kitazawa:2017qab}%
  \BibitemOpen
  \bibfield  {author} {\bibinfo {author} {\bibfnamefont {M.}~\bibnamefont
  {Kitazawa}}, \bibinfo {author} {\bibfnamefont {T.}~\bibnamefont {Iritani}},
  \bibinfo {author} {\bibfnamefont {M.}~\bibnamefont {Asakawa}}, \ and\
  \bibinfo {author} {\bibfnamefont {T.}~\bibnamefont {Hatsuda}},\ }\href
  {\doibase 10.1103/PhysRevD.96.111502} {\bibfield  {journal} {\bibinfo
  {journal} {Phys. Rev. D}\ }\textbf {\bibinfo {volume} {96}},\ \bibinfo
  {pages} {111502} (\bibinfo {year} {2017})}\BibitemShut {NoStop}%
\bibitem [{\citenamefont {Borsanyi}\ \emph {et~al.}(2012)\citenamefont
  {Borsanyi} \emph {et~al.}}]{Borsanyi:2012zs}%
  \BibitemOpen
  \bibfield  {author} {\bibinfo {author} {\bibfnamefont {S.}~\bibnamefont
  {Borsanyi}} \emph {et~al.},\ }\href {\doibase 10.1007/JHEP09(2012)010}
  {\bibfield  {journal} {\bibinfo  {journal} {JHEP}\ }\textbf {\bibinfo
  {volume} {09}},\ \bibinfo {pages} {010} (\bibinfo {year} {2012})}\BibitemShut
  {NoStop}%
\bibitem [{\citenamefont {Bazavov}\ \emph {et~al.}(2016)\citenamefont {Bazavov}
  \emph {et~al.}}]{Bazavov:2015yea}%
  \BibitemOpen
  \bibfield  {author} {\bibinfo {author} {\bibfnamefont {A.}~\bibnamefont
  {Bazavov}} \emph {et~al.} (\bibinfo {collaboration} {MILC}),\ }\href
  {\doibase 10.1103/PhysRevD.93.094510} {\bibfield  {journal} {\bibinfo
  {journal} {Phys. Rev.}\ }\textbf {\bibinfo {volume} {D93}},\ \bibinfo {pages}
  {094510} (\bibinfo {year} {2016})}\BibitemShut {NoStop}%
\bibitem [{\citenamefont {Dalla~Brida}\ and\ \citenamefont
  {Ramos}(2019)}]{DallaBrida:2019wur}%
  \BibitemOpen
  \bibfield  {author} {\bibinfo {author} {\bibfnamefont {M.}~\bibnamefont
  {Dalla~Brida}}\ and\ \bibinfo {author} {\bibfnamefont {A.}~\bibnamefont
  {Ramos}},\ }\href {\doibase 10.1140/epjc/s10052-019-7228-z} {\bibfield
  {journal} {\bibinfo  {journal} {Eur. Phys. J.}\ }\textbf {\bibinfo {volume}
  {C79}},\ \bibinfo {pages} {720} (\bibinfo {year} {2019})}\BibitemShut
  {NoStop}%
\bibitem [{\citenamefont {Cè}\ \emph {et~al.}(2015)\citenamefont {Cè},
  \citenamefont {Consonni}, \citenamefont {Engel},\ and\ \citenamefont
  {Giusti}}]{Ce:2015qha}%
  \BibitemOpen
  \bibfield  {author} {\bibinfo {author} {\bibfnamefont {M.}~\bibnamefont
  {Cè}}, \bibinfo {author} {\bibfnamefont {C.}~\bibnamefont {Consonni}},
  \bibinfo {author} {\bibfnamefont {G.~P.}\ \bibnamefont {Engel}}, \ and\
  \bibinfo {author} {\bibfnamefont {L.}~\bibnamefont {Giusti}},\ }\href
  {\doibase 10.1103/PhysRevD.92.074502} {\bibfield  {journal} {\bibinfo
  {journal} {Phys. Rev.}\ }\textbf {\bibinfo {volume} {D92}},\ \bibinfo {pages}
  {074502} (\bibinfo {year} {2015})}\BibitemShut {NoStop}%
\bibitem [{\citenamefont {Taniguchi}\ \emph {et~al.}(2017)\citenamefont
  {Taniguchi}, \citenamefont {Kanaya}, \citenamefont {Suzuki},\ and\
  \citenamefont {Umeda}}]{Taniguchi:2016tjc}%
  \BibitemOpen
  \bibfield  {author} {\bibinfo {author} {\bibfnamefont {Y.}~\bibnamefont
  {Taniguchi}}, \bibinfo {author} {\bibfnamefont {K.}~\bibnamefont {Kanaya}},
  \bibinfo {author} {\bibfnamefont {H.}~\bibnamefont {Suzuki}}, \ and\ \bibinfo
  {author} {\bibfnamefont {T.}~\bibnamefont {Umeda}},\ }\href {\doibase
  10.1103/PhysRevD.95.054502} {\bibfield  {journal} {\bibinfo  {journal} {Phys.
  Rev.}\ }\textbf {\bibinfo {volume} {D95}},\ \bibinfo {pages} {054502}
  (\bibinfo {year} {2017})}\BibitemShut {NoStop}%
\bibitem [{\citenamefont {Mazur}\ \emph {et~al.}(2020)\citenamefont {Mazur},
  \citenamefont {Altenkort}, \citenamefont {Kaczmarek},\ and\ \citenamefont
  {Shu}}]{Mazur:2020hvt}%
  \BibitemOpen
  \bibfield  {author} {\bibinfo {author} {\bibfnamefont {L.}~\bibnamefont
  {Mazur}}, \bibinfo {author} {\bibfnamefont {L.}~\bibnamefont {Altenkort}},
  \bibinfo {author} {\bibfnamefont {O.}~\bibnamefont {Kaczmarek}}, \ and\
  \bibinfo {author} {\bibfnamefont {H.-T.}\ \bibnamefont {Shu}},\ }\href
  {\doibase 10.22323/1.363.0219} {\bibfield  {journal} {\bibinfo  {journal}
  {PoS}\ }\textbf {\bibinfo {volume} {LATTICE2019}},\ \bibinfo {pages} {219}
  (\bibinfo {year} {2020})}\BibitemShut {NoStop}%
\bibitem [{\citenamefont {Luscher}(2010)}]{Luscher:2010we}%
  \BibitemOpen
  \bibfield  {author} {\bibinfo {author} {\bibfnamefont {M.}~\bibnamefont
  {Luscher}},\ }\href {\doibase 10.22323/1.105.0015} {\bibfield  {journal}
  {\bibinfo  {journal} {PoS}\ }\textbf {\bibinfo {volume} {LATTICE2010}},\
  \bibinfo {pages} {015} (\bibinfo {year} {2010})}\BibitemShut {NoStop}%
\bibitem [{\citenamefont {Altenkort}\ \emph
  {et~al.}(2021{\natexlab{a}})\citenamefont {Altenkort}, \citenamefont {Eller},
  \citenamefont {Kaczmarek}, \citenamefont {Mazur}, \citenamefont {Moore},\
  and\ \citenamefont {Shu}}]{Altenkort:2020fgs}%
  \BibitemOpen
  \bibfield  {author} {\bibinfo {author} {\bibfnamefont {L.}~\bibnamefont
  {Altenkort}}, \bibinfo {author} {\bibfnamefont {A.~M.}\ \bibnamefont
  {Eller}}, \bibinfo {author} {\bibfnamefont {O.}~\bibnamefont {Kaczmarek}},
  \bibinfo {author} {\bibfnamefont {L.}~\bibnamefont {Mazur}}, \bibinfo
  {author} {\bibfnamefont {G.~D.}\ \bibnamefont {Moore}}, \ and\ \bibinfo
  {author} {\bibfnamefont {H.-T.}\ \bibnamefont {Shu}},\ }\href {\doibase
  10.1103/PhysRevD.103.014511} {\bibfield  {journal} {\bibinfo  {journal}
  {Phys. Rev. D}\ }\textbf {\bibinfo {volume} {103}},\ \bibinfo {pages}
  {014511} (\bibinfo {year} {2021}{\natexlab{a}})}\BibitemShut {NoStop}%
\bibitem [{\citenamefont {Francis}\ \emph {et~al.}(2015)\citenamefont
  {Francis}, \citenamefont {Kaczmarek}, \citenamefont {Laine}, \citenamefont
  {Neuhaus},\ and\ \citenamefont {Ohno}}]{Francis:2015lha}%
  \BibitemOpen
  \bibfield  {author} {\bibinfo {author} {\bibfnamefont {A.}~\bibnamefont
  {Francis}}, \bibinfo {author} {\bibfnamefont {O.}~\bibnamefont {Kaczmarek}},
  \bibinfo {author} {\bibfnamefont {M.}~\bibnamefont {Laine}}, \bibinfo
  {author} {\bibfnamefont {T.}~\bibnamefont {Neuhaus}}, \ and\ \bibinfo
  {author} {\bibfnamefont {H.}~\bibnamefont {Ohno}},\ }\href {\doibase
  10.1103/PhysRevD.91.096002} {\bibfield  {journal} {\bibinfo  {journal} {Phys.
  Rev. D}\ }\textbf {\bibinfo {volume} {91}},\ \bibinfo {pages} {096002}
  (\bibinfo {year} {2015})}\BibitemShut {NoStop}%
\bibitem [{\citenamefont {Sommer}(1994)}]{Sommer:1993ce}%
  \BibitemOpen
  \bibfield  {author} {\bibinfo {author} {\bibfnamefont {R.}~\bibnamefont
  {Sommer}},\ }\href {\doibase 10.1016/0550-3213(94)90473-1} {\bibfield
  {journal} {\bibinfo  {journal} {Nucl. Phys. B}\ }\textbf {\bibinfo {volume}
  {411}},\ \bibinfo {pages} {839} (\bibinfo {year} {1994})}\BibitemShut
  {NoStop}%
\bibitem [{\citenamefont {Burnier}\ \emph {et~al.}(2017)\citenamefont
  {Burnier}, \citenamefont {Ding}, \citenamefont {Kaczmarek}, \citenamefont
  {Kruse}, \citenamefont {Laine}, \citenamefont {Ohno},\ and\ \citenamefont
  {Sandmeyer}}]{Burnier:2017bod}%
  \BibitemOpen
  \bibfield  {author} {\bibinfo {author} {\bibfnamefont {Y.}~\bibnamefont
  {Burnier}}, \bibinfo {author} {\bibfnamefont {H.~T.}\ \bibnamefont {Ding}},
  \bibinfo {author} {\bibfnamefont {O.}~\bibnamefont {Kaczmarek}}, \bibinfo
  {author} {\bibfnamefont {A.~L.}\ \bibnamefont {Kruse}}, \bibinfo {author}
  {\bibfnamefont {M.}~\bibnamefont {Laine}}, \bibinfo {author} {\bibfnamefont
  {H.}~\bibnamefont {Ohno}}, \ and\ \bibinfo {author} {\bibfnamefont
  {H.}~\bibnamefont {Sandmeyer}},\ }\href {\doibase 10.1007/JHEP11(2017)206}
  {\bibfield  {journal} {\bibinfo  {journal} {JHEP}\ }\textbf {\bibinfo
  {volume} {11}},\ \bibinfo {pages} {206} (\bibinfo {year} {2017})}\BibitemShut
  {NoStop}%
\bibitem [{\citenamefont {Bilson-Thompson}\ \emph {et~al.}(2003)\citenamefont
  {Bilson-Thompson}, \citenamefont {Leinweber},\ and\ \citenamefont
  {Williams}}]{BilsonThompson:2002jk}%
  \BibitemOpen
  \bibfield  {author} {\bibinfo {author} {\bibfnamefont {S.~O.}\ \bibnamefont
  {Bilson-Thompson}}, \bibinfo {author} {\bibfnamefont {D.~B.}\ \bibnamefont
  {Leinweber}}, \ and\ \bibinfo {author} {\bibfnamefont {A.~G.}\ \bibnamefont
  {Williams}},\ }\href {\doibase 10.1016/S0003-4916(03)00009-5} {\bibfield
  {journal} {\bibinfo  {journal} {Annals Phys.}\ }\textbf {\bibinfo {volume}
  {304}},\ \bibinfo {pages} {1} (\bibinfo {year} {2003})}\BibitemShut {NoStop}%
\bibitem [{\citenamefont {Ramos}\ and\ \citenamefont
  {Sint}(2016)}]{Ramos:2015baa}%
  \BibitemOpen
  \bibfield  {author} {\bibinfo {author} {\bibfnamefont {A.}~\bibnamefont
  {Ramos}}\ and\ \bibinfo {author} {\bibfnamefont {S.}~\bibnamefont {Sint}},\
  }\href {\doibase 10.1140/epjc/s10052-015-3831-9} {\bibfield  {journal}
  {\bibinfo  {journal} {Eur. Phys. J. C}\ }\textbf {\bibinfo {volume} {76}},\
  \bibinfo {pages} {15} (\bibinfo {year} {2016})}\BibitemShut {NoStop}%
\bibitem [{\citenamefont {Eller}\ and\ \citenamefont
  {Moore}(2018)}]{Eller:2018yje}%
  \BibitemOpen
  \bibfield  {author} {\bibinfo {author} {\bibfnamefont {A.~M.}\ \bibnamefont
  {Eller}}\ and\ \bibinfo {author} {\bibfnamefont {G.~D.}\ \bibnamefont
  {Moore}},\ }\href {\doibase 10.1103/PhysRevD.97.114507} {\bibfield  {journal}
  {\bibinfo  {journal} {Phys. Rev.}\ }\textbf {\bibinfo {volume} {D97}},\
  \bibinfo {pages} {114507} (\bibinfo {year} {2018})}\BibitemShut {NoStop}%
\bibitem [{\citenamefont {Arnold}\ and\ \citenamefont
  {McLerran}(1988)}]{Arnold:1987zg}%
  \BibitemOpen
  \bibfield  {author} {\bibinfo {author} {\bibfnamefont {P.~B.}\ \bibnamefont
  {Arnold}}\ and\ \bibinfo {author} {\bibfnamefont {L.~D.}\ \bibnamefont
  {McLerran}},\ }\href {\doibase 10.1103/PhysRevD.37.1020} {\bibfield
  {journal} {\bibinfo  {journal} {Phys. Rev. D}\ }\textbf {\bibinfo {volume}
  {37}},\ \bibinfo {pages} {1020} (\bibinfo {year} {1988})}\BibitemShut
  {NoStop}%
\bibitem [{\citenamefont {Suzuki}\ and\ \citenamefont
  {Takaura}(2021)}]{Suzuki:2021tlr}%
  \BibitemOpen
  \bibfield  {author} {\bibinfo {author} {\bibfnamefont {H.}~\bibnamefont
  {Suzuki}}\ and\ \bibinfo {author} {\bibfnamefont {H.}~\bibnamefont
  {Takaura}},\ }\href@noop {} {\  (\bibinfo {year} {2021})},\ \bibinfo {note}
  {arXiv:\href{https://arxiv.org/abs/2102.02174}{2102.02174}}\BibitemShut
  {NoStop}%
\bibitem [{\citenamefont {Seiler}(2002)}]{Seiler:2001je}%
  \BibitemOpen
  \bibfield  {author} {\bibinfo {author} {\bibfnamefont {E.}~\bibnamefont
  {Seiler}},\ }\href {\doibase 10.1016/S0370-2693(01)01469-1} {\bibfield
  {journal} {\bibinfo  {journal} {Phys. Lett. B}\ }\textbf {\bibinfo {volume}
  {525}},\ \bibinfo {pages} {355} (\bibinfo {year} {2002})}\BibitemShut
  {NoStop}%
\bibitem [{\citenamefont {Vicari}(1999)}]{Vicari:1999xx}%
  \BibitemOpen
  \bibfield  {author} {\bibinfo {author} {\bibfnamefont {E.}~\bibnamefont
  {Vicari}},\ }\href {\doibase 10.1016/S0550-3213(99)00297-7} {\bibfield
  {journal} {\bibinfo  {journal} {Nucl. Phys. B}\ }\textbf {\bibinfo {volume}
  {554}},\ \bibinfo {pages} {301} (\bibinfo {year} {1999})}\BibitemShut
  {NoStop}%
\bibitem [{\citenamefont {Horvath}\ \emph {et~al.}(2005)\citenamefont
  {Horvath}, \citenamefont {Alexandru}, \citenamefont {Zhang}, \citenamefont
  {Chen}, \citenamefont {Dong}, \citenamefont {Draper}, \citenamefont {Liu},
  \citenamefont {Mathur}, \citenamefont {Tamhankar},\ and\ \citenamefont
  {Thacker}}]{Horvath:2005cv}%
  \BibitemOpen
  \bibfield  {author} {\bibinfo {author} {\bibfnamefont {I.}~\bibnamefont
  {Horvath}}, \bibinfo {author} {\bibfnamefont {A.}~\bibnamefont {Alexandru}},
  \bibinfo {author} {\bibfnamefont {J.}~\bibnamefont {Zhang}}, \bibinfo
  {author} {\bibfnamefont {Y.}~\bibnamefont {Chen}}, \bibinfo {author}
  {\bibfnamefont {S.}~\bibnamefont {Dong}}, \bibinfo {author} {\bibfnamefont
  {T.}~\bibnamefont {Draper}}, \bibinfo {author} {\bibfnamefont
  {K.}~\bibnamefont {Liu}}, \bibinfo {author} {\bibfnamefont {N.}~\bibnamefont
  {Mathur}}, \bibinfo {author} {\bibfnamefont {S.}~\bibnamefont {Tamhankar}}, \
  and\ \bibinfo {author} {\bibfnamefont {H.}~\bibnamefont {Thacker}},\ }\href
  {\doibase 10.1016/j.physletb.2005.04.076} {\bibfield  {journal} {\bibinfo
  {journal} {Phys. Lett. B}\ }\textbf {\bibinfo {volume} {617}},\ \bibinfo
  {pages} {49} (\bibinfo {year} {2005})}\BibitemShut {NoStop}%
\bibitem [{\citenamefont {Laine}\ \emph {et~al.}(2011)\citenamefont {Laine},
  \citenamefont {Vuorinen},\ and\ \citenamefont {Zhu}}]{Laine:2011xm}%
  \BibitemOpen
  \bibfield  {author} {\bibinfo {author} {\bibfnamefont {M.}~\bibnamefont
  {Laine}}, \bibinfo {author} {\bibfnamefont {A.}~\bibnamefont {Vuorinen}}, \
  and\ \bibinfo {author} {\bibfnamefont {Y.}~\bibnamefont {Zhu}},\ }\href
  {\doibase 10.1007/JHEP09(2011)084} {\bibfield  {journal} {\bibinfo  {journal}
  {JHEP}\ }\textbf {\bibinfo {volume} {09}},\ \bibinfo {pages} {084} (\bibinfo
  {year} {2011})}\BibitemShut {NoStop}%
\bibitem [{\citenamefont {Arnold}\ \emph {et~al.}(1997)\citenamefont {Arnold},
  \citenamefont {Son},\ and\ \citenamefont {Yaffe}}]{Arnold:1996dy}%
  \BibitemOpen
  \bibfield  {author} {\bibinfo {author} {\bibfnamefont {P.~B.}\ \bibnamefont
  {Arnold}}, \bibinfo {author} {\bibfnamefont {D.}~\bibnamefont {Son}}, \ and\
  \bibinfo {author} {\bibfnamefont {L.~G.}\ \bibnamefont {Yaffe}},\ }\href
  {\doibase 10.1103/PhysRevD.55.6264} {\bibfield  {journal} {\bibinfo
  {journal} {Phys. Rev. D}\ }\textbf {\bibinfo {volume} {55}},\ \bibinfo
  {pages} {6264} (\bibinfo {year} {1997})}\BibitemShut {NoStop}%
\bibitem [{\citenamefont {Altenkort}\ \emph
  {et~al.}(2021{\natexlab{b}})\citenamefont {Altenkort}, \citenamefont {Eller},
  \citenamefont {Kaczmarek}, \citenamefont {Mazur}, \citenamefont {Moore},\
  and\ \citenamefont {Shu}}]{datapublication}%
  \BibitemOpen
  \bibfield  {author} {\bibinfo {author} {\bibfnamefont {L.}~\bibnamefont
  {Altenkort}}, \bibinfo {author} {\bibfnamefont {A.~M.}\ \bibnamefont
  {Eller}}, \bibinfo {author} {\bibfnamefont {O.}~\bibnamefont {Kaczmarek}},
  \bibinfo {author} {\bibfnamefont {L.}~\bibnamefont {Mazur}}, \bibinfo
  {author} {\bibfnamefont {G.~D.}\ \bibnamefont {Moore}}, \ and\ \bibinfo
  {author} {\bibfnamefont {H.-T.}\ \bibnamefont {Shu}},\ }\href {\doibase
  10.4119/unibi/2954712} {\bibfield  {journal} {\bibinfo  {journal} {Bielefeld
  University}\ } (\bibinfo {year} {2021}{\natexlab{b}}),\
  10.4119/unibi/2954712}\BibitemShut {NoStop}%
\end{thebibliography}%

\end{document}